\documentclass[twocolumn]{aa} 
\usepackage{graphicx}
\usepackage{txfonts}
\usepackage{natbib}
\begin{document}
   \title{NACO-SDI imaging of known companion host stars from the AAPS and Keck planet search surveys}


   \author{J.S.~Jenkins\inst{1}
	  \and 
	  H.R.A.~Jones\inst{2}
	  \and 
	  B.~Biller\inst{3}
          \and
          S.J.~O'Toole\inst{4}
	  \and 
	  D.J.~Pinfield\inst{2}
	  \and 
	  L.~Close\inst{5}
	  \and 
	  C.G.~Tinney\inst{6}
	  \and 
	  R.P.~Butler\inst{7}
	  \and 
	  R.~Wittenmyer\inst{6}
	  \and 
	  B.~Carter\inst{8}
          \and
          A.C.~Day-Jones\inst{1}
   }

   \offprints{J.S Jenkins}

   \institute{$^1$Department of Astronomy, Universidad de Chile, Casilla 36-D, Santiago, Chile \\
$^2$Center for Astrophysics, University of Hertfordshire, College Lane Campus, Hatfield, Hertfordshire, UK, AL10 9AB\\
$^3$Institute of Astronomy, 2680 Woodlawn Drive, Honolulu, HI 96822 \\
$^4$Anglo-Australian Observatory, PO Box 296, Epping 1710, Australia \\
$^5$Steward Observatory, University of Arizona, Tucson, AZ 85721 \\
$^6$Department of Astrophysics \& Optics, University of New South Wales, NSW 2052, Australia \\
$^7$Carnegie Institute of Washington, Department of Terrestrial Magnetism, 5241 Broad Branch Road NW, Washington, DC 20015-1305 \\  
$^{8}$Faculty of Sciences, University of Southern Queensland, Toowoomba, 4350, Australia \\
              \email{jjenkins@das.uchile.cl}
	      \thanks{Based on observations made with the ESO telescopes at the La Silla Paranal observatory under program ID 076.C-0877(B)}
            }

   \date{Received September 2nd 2007}

 
  \abstract
   {Direct imaging of brown dwarfs as companions to solar-type stars can provide a wealth of well-constrained data to ``benchmark''
the physics of such objects, since quantities like metallicity and age can be determined from their well-studied primaries.}
   {We present results from an adaptive optics imaging program on stars drawn from the Anglo-Australian and Keck Planet Search projects,
with the aim of directly imaging known cool companions.    
}
   {Simulations have modeled the expected contrast ratios and separations of known companions using estimates of orbital parameters available from current radial-velocity 
data and then a selection of the best case objects were followed-up with high contrast imaging to attempt to directly image these companions.
}
   {These simulations suggest that only a very small number of radial-velocity detected exoplanets with consistent velocity fits and age estimates could potentially be directly 
imaged using the VLT's Simultaneous Differential Imaging system and only under favorable conditions.  We also present detectability confidence limits 
from the radial-velocity data sets and show how these can be used to gain a better understanding of these systems when combined with the imaging data.  

For HD32778 and HD91204 the detectabilities help little in constraining the 
companion and hence almost all our knowledge is drawn from the SDI images.  Therefore, we can say that these stars do not host cool methane objects, out to on-sky separations 
of $\sim$2$''$, with contrasts less than 10--11~magnitudes.  However, for HD25874, HD120780 and HD145825, the contrasts and detectabilities can rule out a number of 
possible solutions, particularly at low angular separations, and for the best case, down to strong methane masses of 40M$_{\rm{J}}$ at 1$''$ separation.  The contrast curves 
constructed for these five stars show 5$\sigma$ contrasts ($\Delta$F1) of $\sim$9.2--11.5 magnitudes at separations of $\ge$0.6$''$, which correspond to contrasts of
$\sim$9.7--12.0 magnitudes for companions of mid-T spectral type.  Such limits allow us to reach down to 40M$_{\rm{J}}$ around fairly old field dwarfs that typically 
constitute high precision radial-velocity programs.  Finally, the analysis performed here can serve as a template for future projects that will employ extreme-AO 
systems to directly image planets already indirectly discovered by the radial-velocity method.
}
   {}

   \keywords{Stars: brown dwarfs --
		Stars: planetary systems
               }

   \maketitle
%

\section{Introduction}

The detection of over 400 planets orbiting Sun-like stars has revolutionised our knowledge of our local neighbourhood
and our position therein.  Yet planets are not the sole close companions to solar-type stars.  For instance,
\citet{duquennoy,duquennoy91} and \citet{duquennoy92} have examined stellar multiplicity in a series of papers.  Radial-velocity surveys have revealed few brown dwarfs 
orbiting solar-type stars (e.g. \citealp{wittenmyer09}; \citealp{jenkins09a}) leading to the phrase `brown dwarf desert' being 
coined to describe this paucity (\citealp{marcy}).  However, beyond $\sim$4AU one would expect few radial-velocity planetary or brown dwarf companions to be known due to the 
limited temporal coverage at the required precision levels necessary to fully sample such companions.  In addition, radial-velocity surveys also have strong biases against the detection 
of long-period companions, as the radial-velocity amplitude is a strong function 
of orbital period and also since this technique requires the observation of at least half an orbit (e.g. \citealp{wright07}) to constrain companion properties.  Only now are we sensitive enough 
to detect solar system-like gas giant planets in solar system-like orbits (e.g. \citealp{jones10}).

Conversely, direct and coronographic imaging techniques can probe much wider separations than current radial-velocity programs can reach.  
For example, \citet{kalas08} and \citet{marois08} have directly imaged planetary mass companions to the stars Fomalhaut and HR~8799, located at angular separations of 14.9$''$ and 1.73$''$, or 115AU and 68AU, respectively.  \citet{mccarthy04} found another deficit of brown dwarf companions between 75-1200 AU.  \citet{liu02} used the Gemini-North
and Keck Adaptive Optics (AO) systems to obtain three epochs of images of the brown dwarf companion to HR~7672, which had initially been detected by its radial-velocity
signature.  The flux ratio at 2.16$\mu$m was found to be 8.6 magnitudes at a separation of 0.79$''$.  This level of contrast pushed the instrumentation used in this detection to its 
very limits.  However the introduction of Simultaneous
Differential Imaging (SDI) on the VLT's NACO facility permits the achievement of higher contrasts, at smaller separations, for the coolest stellar companions.  For example, 
contrasts on the order of $\Delta$H$\sim$13 have been demonstrated at $\sim$0.5$''$ by \citet{mugrauer} and \citet{biller07}.


\section{Target Selection}

In order to guide the selection of target host stars for adaptive optics imaging of brown dwarfs and exoplanets, we have performed simulations which take the best currently available 
companion parameters from radial-velocity data sets, combined with host-star age estimates and brown dwarf and exoplanetary interior models, to derive predicted magnitude differences 
and angular separations on sky.  These simulations were performed for all stars in the Anglo-Australian and Keck Planet Searches (for samples see \citealp{jones02a}, \citealp{marcy05a}, 
\citealp{butler06} and references therein), which show a long term radial-velocity profile consistent with an orbiting low-mass companion.

\subsection{Angular Separation}

Hipparcos distance data (\citealp{vanleeuwen05}) is available for all these objects (which all lie at distances of less than 100pc).  It should be noted that in most 
cases, the radial-velocity orbital solutions are not well constrained.  This is largely because the companion orbits are much longer than the monitoring baselines of the surveys, and in some 
cases because the companion properties have been derived with no inflection in the radial-velocity curve (often referred to in the planet searches as a ``liner''). The fits to both 
these classes of data produce only semi-major axis lower limits.  In addition, the eccentricities of most of the companions are so poorly constrained that they are fixed to zero, 
causing further separation ambiguity.

\begin{figure}
\vspace{5.5cm}
\includegraphics{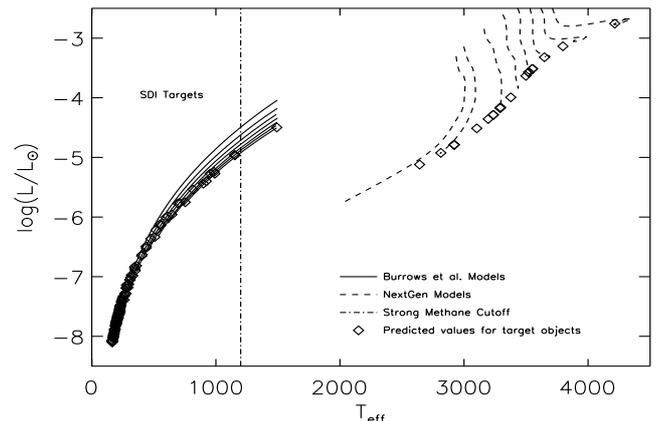}
\vspace{0.6cm}
\caption[HR-diagram with model fits to all the cool dwarf candidates]{The solid lines are \citet{burrows} models
for different masses and ages.  The dashed lines represent a range of masses and
ages for the NextGen models of \citet{allard}, which are populated by our high-mass targets.  The dot-dashed line
marks the methane absorption boundary in cool dwarfs and provides a useful upper limit for SDI targets.  The discontinuity between the models is the L to T spectral type boundary
region, which is not very well modeled, thus the Burrows et al. models were truncated.}
\label{hr_diagram}
\end{figure}

\subsection{Contrast Ratio}

Infrared photometry for the primary stars were taken from the 2MASS catalogue (http://irsa.-ipac.caltech.edu/) and when combined with Hipparcos
distances we were able to generate accurate absolute JHK$_{s}$ magnitudes.  The K$_{s}$ from 2MASS was converted to K using the magnitude corrections in \citet{carpenter01}.  Absolute 
magnitudes for the companions were estimated using the non-grey evolutionary tracks of \citet{burrows}, 
the COND models (\citealp{baraffe03}) and the NextGen models of \citet{allard}.  The masses of these simulated candidate companions were taken from the radial-velocity data and range 
from 1-725M$_{\rm{J}}$.  Ages for the systems were taken from \citet{valenti05} and \citet{takeda07}, which limits the size of our sample but retains a high level of internal consistency.  

We split the companions into two groups depending on which model 
we could use to generate robust magnitudes.  The split was made on the basis of companion \emph{T}$_{\rm{EFF}}$ values, with all the companions using the Burrows et al.
models having a \emph{T}$_{\rm{EFF}}$~$<$~2000K (which for these older systems corresponds to a M~sin~\emph{i}~$<$~85M$_{\rm{J}}$) and all other companions using the NextGen models.  A custom spline
fitting procedure obtained the bolometric luminosities for all the companions (see Fig.~\ref{hr_diagram}).  The separation in effective temperature between the low temperature models 
of Burrows et al. and Baraffe et al. and the higher temperature models of Allard et al. is clearly apparent.  Taking the whole \citet{butler06} catalogue there are 156 companions below 
the strong methane absorption boundary (1200K) (shown by the dot-dashed
vertical line on the plot), apart from their low luminosity we expect them to be good SDI targets and a small number of these may well be detectable.  There are
also 23 companions with \emph{T}$_{\rm{EFF}}$ values above 1200K, 22 of which have T$_{\rm{eff}}~>~$2000K.  These should be observable with conventional AO methods.

\begin{figure}
\vspace{4.5cm}
\includegraphics{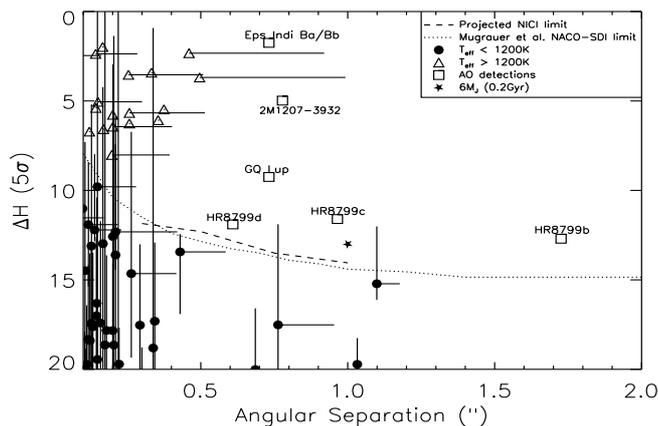}
\vspace{1.35cm}
\caption[Contrast ratio as a function of separation for the cool dwarf candidates]{A subset of our simulated candidate companions, including those imaged in this paper, taken from 
Butler et al. truncated by $\Delta$H.  Triangles represent targets with T$_{\rm{eff}}~>$~1200K,
filled circles are targets $<$~1200K (potentially amenable to SDI) and the squares are companions to young stars already found by high contrast AO imaging.  For comparison,
the dotted line shows the NACO-SDI sensitivity (\citealp{mugrauer}) and the dashed line shows the NICI sensitivity taken from the Gemini website
(http://www.gemini.edu/sciops/instruments/nici).
The error bars represent the range of possible ages from \cite{valenti05} and \cite{takeda07}, along with the error in the RV measurements.  The star represents a 6M$_{\rm{J}}$ 
object at 5AU orbiting a typical K0 star with an age of 0.2~Gyr and a distance of 5pc.}
\label{cont_sep}
\end{figure}

The major problem faced when attempting to image close-in companions to bright stars, is the contrast difference.  To determine the contrasts between the
stars and their companions we had to obtain the JHK magnitudes of the companions, both from internal and reflected flux.  To accomplish this the median colors and
bolometric corrections (BC) from \citet{leggett} (BC$_{\rm{T}}$=2.06;~BC$_{\rm{L}}$=3.25, the subscripts denote spectral type) were employed.  This correction gave the
absolute K magnitudes and, using the colors (H-K$_{\rm{T}}$=-0.04;~H-K$_{\rm{L}}$=0.70), we were able to generate their expected absolute H-band magnitudes.  For all
planetary-mass companions Jupiter characteristics were assumed.  We then
simplistically simulated the expected reflected flux in the $H$-band and added this component to the internal flux.
This was done by using simple geometry and assuming Jupiter's radius and albedo.  The total absolute magnitudes
of the companions were then subtracted from the absolute magnitudes derived for the primaries giving the estimated contrast $\Delta$H.

Figure~\ref{cont_sep} shows
the contrast and separation estimates from the simulation for all the radial-velocity companions included in this test.  The simulated companions represented by filled circles 
would all have strong methane absorption.  The bulk of the objects lie within 0.3$''$ of the primaries due to the bias introduced by the short radial-velocity baseline.  The
majority of these lower-mass companions (M~sin~\emph{i}~$<$~15M$_{\rm{J}}$) also have high contrasts
($\Delta$H$>>$10), putting them below the 5$\sigma$ NACO-SDI threshold.  However, two companions have larger separations
$>$0.3$''$, approaching the separations of the already discovered objects of \citet{marois08}, \citet{mccaughrean03}, \citet{chauvin} and \citet{neuhauser} and, combined with a
$\Delta$H$<$13.5, they could be amenable to SDI imaging.  Note that another secure AO detection is Fomalhaut~$b$ but this is located far off the plot scale with an 
angular separation of $\sim$14.9$''$. 

The NACO-SDI (\citealp{mugrauer}) and NICI (http://www.gemini.edu/sciops/instruments/nici)
sensitivities are highlighted on Fig.~\ref{cont_sep} by dotted and dashed lines respectively.  Note that we can not be sure if the Mugrauer \& Neuhauser detectability limits 
are actual 5$\sigma$ limits or some lower threshold limit.  
Once the masses and semimajor axes are more precisely defined, the companion magnitudes and separations will most likely increase giving lower contrasts
and more viable targets.  This has been highlighted on the plot by the error bars which represent the direction
in which all companions are expected to move once inclination and eccentricity effects are considered and more RV data points acquired.  Another major source of
uncertainty is age.  For example, a typical 1$\sigma$ age uncertainty for these types of objects is $\sim$$\pm$2~Gyr, which translates to a $\sim$$\pm$2~magnitude error
in $\Delta$H with the primary.  Due to the high contrast ratios 
and extremely small separations the majority of these companions are out of reach of current instruments.  However, future Extreme-AO systems which are proposing to reach 
$>$15~magnitudes of contrast may be able to bridge this gap.  

All companions with T$_{\rm{EFF}}~>~$2000K (triangles in Fig.~\ref{cont_sep} and taken from \citealp{nidever} and \citealp{jenkins10}) have $H$-band magnitudes less than 15, allowing 
direct imaging using normal AO
techniques.  Four of these objects have separations larger than 0.35$''$ and $\Delta$H less than 8, making excellent coronographic targets.  All planetary-mass
companions are off the Fig.~\ref{cont_sep} plot scale since they have much larger $H$-band contrasts.  The star in this figure 
shows the position of a 6M$_{\rm{J}}$ planet in a Jupiter-like orbit as a companion to a 0.2~Gyr, K0 star at 5pc.  The age and spectral type were chosen since they relate to the best 
case scenario for one of our objects HD120780.  It shows that by adopting the lower limit to the large errors on the age of this system that the potential exists to detect 
planetary-mass objects around such stars.  Even still, these types of objects reside
extremely close to the plotted instrument thresholds, highlighting just how difficult it is to obtain a direct image of any planetary-mass object with the current suite
of instruments available.  However, radial-velocity studies have revealed a high number ($\ge$28\% of planet hosting stars) of multiple planet systems (\citealp{wright09}), therefore 
imaging planet-host stars can provide useful constraints on any longer period, massive companions not yet revealed in the radial-velocity dataset (e.g. \citealp{mugrauer06}; 
\citealp{mugrauer07}).

\section{Candidate Characteristics}

All radial-velocity data in this section were generated using the AAPS and Keck pipelines.  These pipelines are still undergoing development 
following the procedures and techniques described in \citet{marcy92} and \citet{butler,butler01,butler06}.  The Keplerian fits shown in 
Figs.~\ref{rv_25874},~\ref{rv_32778},~\ref{rv_91204},~\ref{rv_120780}~and~\ref{rv_145825} are 
performed using the Systemic algorithm (\citealp{meschiari09}), however we note that most are not very well constrained using the current radial-velocity data.  
Table~\ref{tab:stars} lists some relevant information for each object relating to both the radial-velocity and 
photometric analysis in this work.  The parameters and their analysis methods can be found in \citet{vanleeuwen05}, \citet{henry}, \citet{valenti05}, \citet{wright05}, \citet{jenkins06c}, \citet{takeda07} and \citet{jenkins08}.  Tables~\ref{tab:rv1},~\ref{tab:rv2},~\ref{tab:rv3},~\ref{tab:rv4}~and~~\ref{tab:rv5} list all radial-velocity data.  

\subsection{HD25874}

\begin{figure}
\vspace{4.5cm}
\hspace{-4.0cm}
\includegraphics{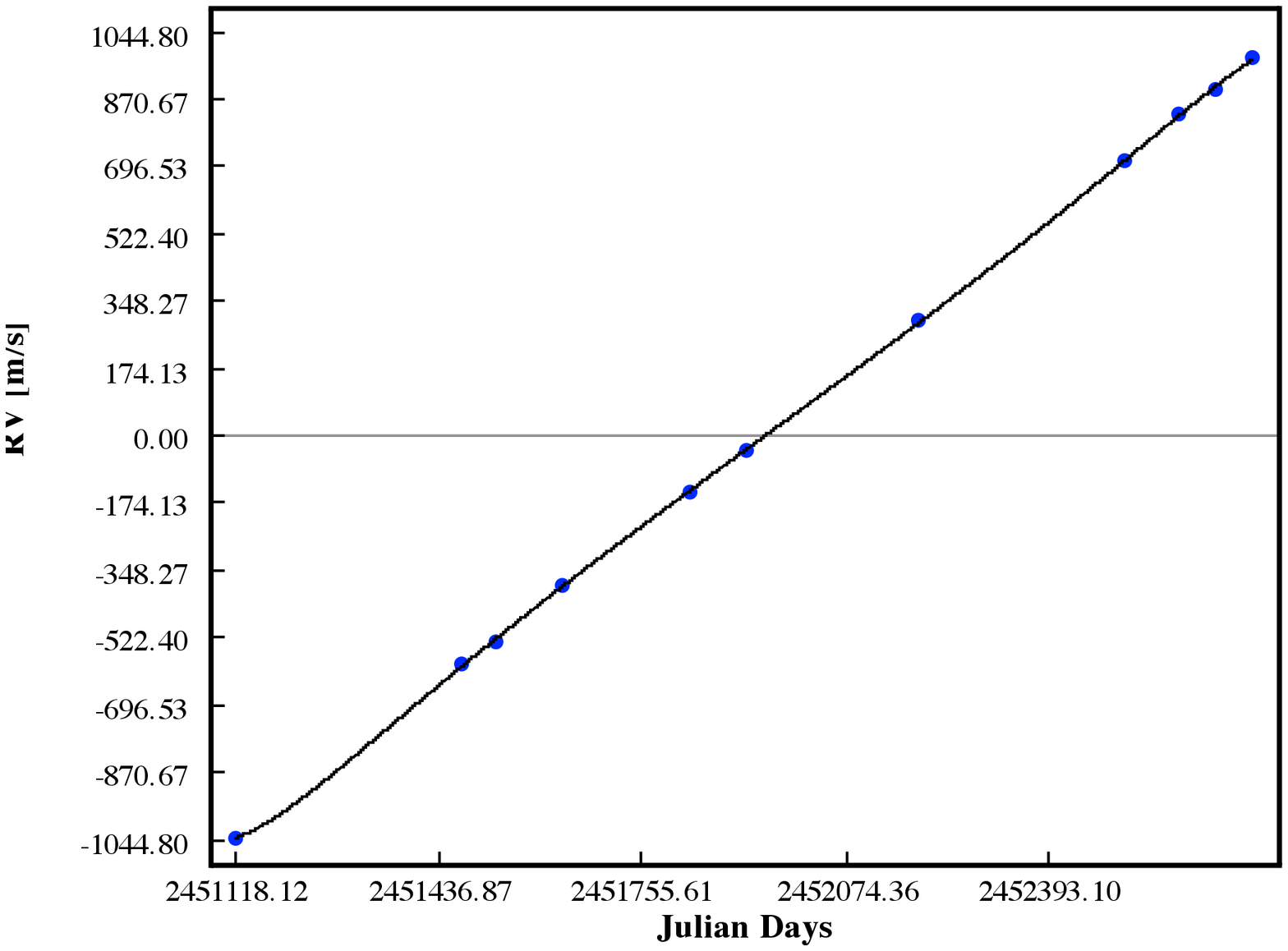}
\vspace{0.5cm}
\caption[RV measurements for HD25874]{The radial-velocity measurements for the star HD25874.  The minimum-period Keplerian fit to the data yields a period of $\sim$6.5 years,
with a companion minimum mass of
66M$_{\rm{J}}$.  However, no curvature has been measured therefore the real orbit will be significantly larger than the orbital fit measured here.}
\label{rv_25874}
\end{figure}

\begin{table*}
\caption[HD25874 data]{Parameters for all stars in this work.} \label{tab:stars}
\center
\begin{tabular}{cccccccccc}
\hline
\multicolumn{1}{c}{Star}& \multicolumn{1}{c}{$V$ (mags)}& \multicolumn{1}{c}{d~(pc)}& \multicolumn{1}{c}{T$_{\rm{EFF}}$ (K)} & \multicolumn{1}{c}{Mass (M$_{\rm{\odot}}$)} & \multicolumn{1}{c}{Radius (R$_{\rm{\odot}}$)}& \multicolumn{1}{c}{Age (Gyrs)}& \multicolumn{1}{c}{[Fe/H]}& \multicolumn{1}{c}{log\emph{R}$'_{\rm{HK}}$}  \\ \hline
& & & & & & \\

HD25874 & 6.74 & 25.91 & 5699 & 1.14$\pm$0.09 & 1.10$\pm$0.03 & 9.36 & -0.02 & -4.95 \\
HD32778 & 7.02 & 22.48 & 5652 & 0.95$\pm$0.08 & 0.86$\pm$0.01 & 10.30 & -0.48 & -4.87 \\
HD91204 & 7.82 & 51.55 & 5914 & 1.05$\pm$0.18 & 1.22$\pm$0.06 & 5.16 & +0.17 & -5.09 \\
HD120780 & 7.37 & 17.01 & 5008 & 0.60$\pm$0.05 & 0.70$\pm$0.01 & 5.40 & -0.26 & -4.79 \\
HD145825 & 6.55 & 21.55 & 5803 & 1.08$\pm$0.10 & 0.97$\pm$0.02 & 1.92 & +0.03 & -4.74 \\

\hline
\end{tabular}
\medskip
\end{table*}

\begin{table}
\caption{HD25874 Radial-velocity data} \label{tab:rv1}
\center
\begin{tabular}{ccc}
\hline
\multicolumn{1}{c}{JD}& \multicolumn{1}{c}{RV~(m/s)}& \multicolumn{1}{c}{$\sigma$$_{\rm{rv}}$~(m/s)} \\ \hline
& & \\

2451118.122  &   -904.1  &    3.4 \\
2451473.261  &   -453.1  &    4.5 \\
2451526.013  &   -394.2  &    3.4 \\
2451630.876  &   -252.7  &    3.4 \\
2451830.118  &   -5.9  &    4.3 \\
2451920.038  &    99.6  &    4.3 \\
2452189.177  &    436.6  &    5.1 \\
2452511.239  &    850.3  &    8.4 \\
2452594.081  &    968.5  &    4.6 \\
2452654.062  &    1032.4  &    4.1 \\
2452710.892  &    1113.6  &    3.2 \\

\hline
\end{tabular}
\medskip
\end{table}

The AAPS has obtained 11 radial-velocity data points over a period of 4.4 years (Fig.~\ref{rv_25874} data taken from \citealp{jenkins10}).  The minimum best-fit Keplerian orbit to 
this data has an amplitude of
$>$1000ms$^{-1}$ relating to a companion period of 6.5~years, eccentricity 0.43 and a minimum mass of 66M$_{\rm{J}}$.  However, the curvature of the fit has 
been generated by the algorithm itself as within the uncertainties all the data points lie in a straight line, known as a liner.  Therefore, the orbital
solutions to this data series are lower limits.  For comparison the best-fit Keplerian with twice the orbital period would relate to a companion minimum mass of $\sim$190M$_{\rm{J}}$ 
and similar $\chi$$^{2}$ of 2.5.  From experience we estimate the lower limit of the period of the orbit to be around four times larger than
currently estimated.  If we take the period range 6.4-25.8~years and the Hipparcos distance of 25.91pc, the projected separation will be in the range 0.13-0.34$''$.  The
absolute H-band magnitude is 3.20 magnitudes and our estimation for the absolute H-band magnitude of the companion using
the technique in Section~3.1 is 15.31 magnitudes, giving a best estimate for the contrast ratio upper limit of 12.11~magnitudes.

\subsection{HD32778}

\begin{table}
\caption{HD32778 Radial-velocity data} \label{tab:rv2}
\center
\begin{tabular}{ccc}
\hline
\multicolumn{1}{c}{JD}& \multicolumn{1}{c}{RV~(m/s)}& \multicolumn{1}{c}{$\sigma$$_{\rm{rv}}$~(m/s)} \\ \hline
& & \\

2452594.134 &  -841.5 &  2.0 \\
2452744.876 &  -502.5 &  2.1 \\
2453042.025 &     0.0 &  1.8 \\
2453046.980 &    19.8 &  2.8 \\
2453402.991 &   448.1 &  2.0 \\

\hline
\end{tabular}
\medskip
\end{table}

\begin{figure}
\vspace{3.5cm}
\hspace{-4.0cm}
\includegraphics{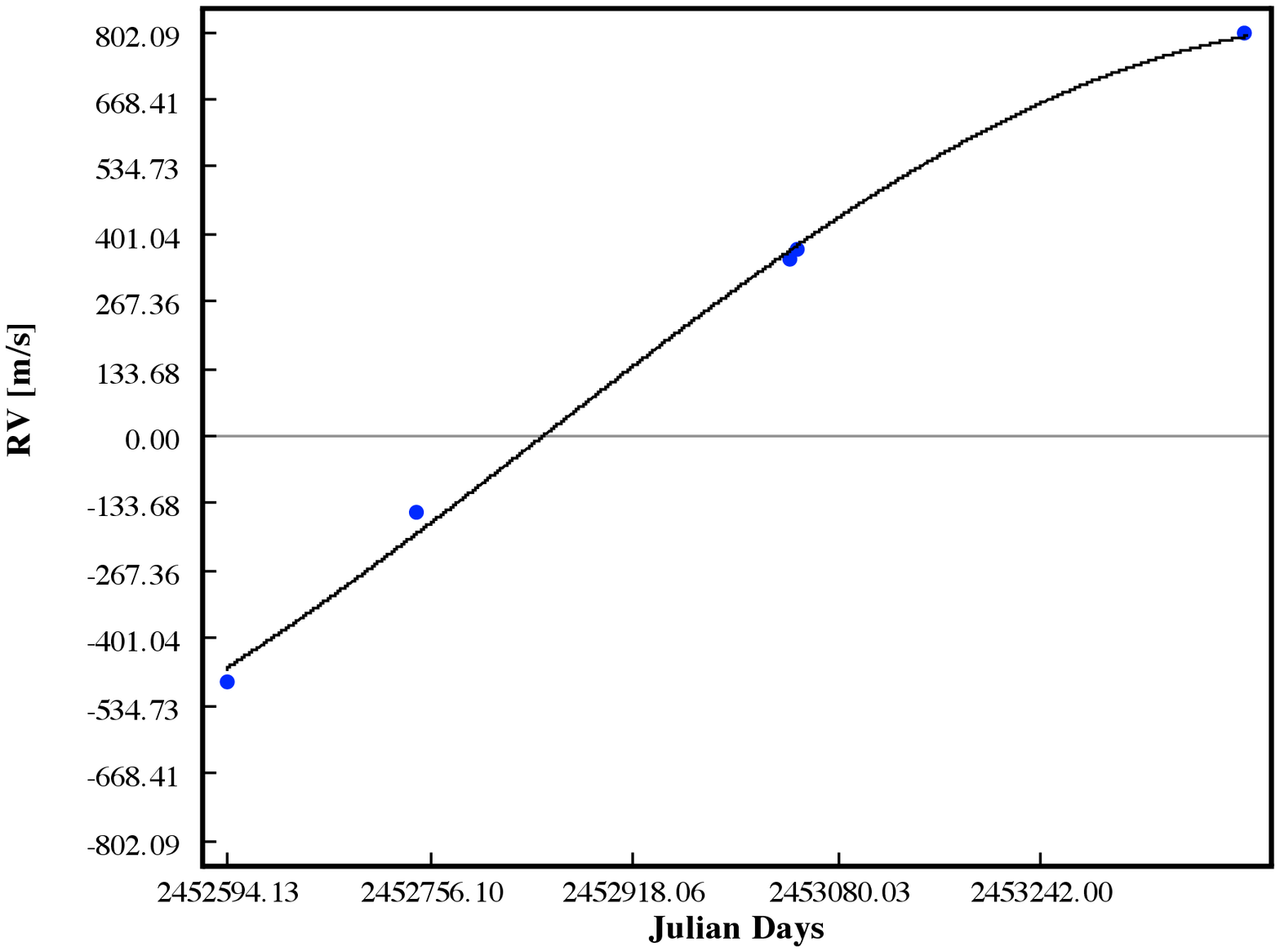}
\vspace{1.cm}
\caption[RV measurements of HD32778]{The best-fit Keplerian orbit to the radial-velocity measurements for the star HD32778, taken by the Keck Planet Search.  The fitted 
orbital period is 6.5~years, with a circular eccentricity and a companion minimum mass of $\sim$55M$_{\rm{J}}$.  Curvature can be seen in this plot, however with only this limited 
number of data points and time coverage the fit is still relatively unconstrained.}
\label{rv_32778}
\end{figure}

Five radial-velocities over a period of 2.25~years for this object (Fig.~\ref{rv_32778}) and the best-fit Keplerian orbit has a semi-amplitude of $\sim$750ms$^{-1}$.
This is consistent with a companion with a minimum mass of $\sim$55M$_{\rm{J}}$, a period of 6.5~years and a circular eccentricity.  Even though this is not a liner the limited amount 
of data points and temporal coverage means this is not very well constrained.  The small amount of curvature does help better constrain the orbit since if we look at the best-fit for 
twice the orbital period we quote here, we find a minimum mass of 137M$_{\rm{J}}$ but with a very high $\chi$$^{2}$ of 186, showing such large orbits are difficult to fit well.  
At a distance of 22.48pc the estimated angular separation for this companion is 0.16$''$.  The absolute 
H-band magnitude of this star is 3.71 and with an estimated maximum absolute H of 15.59 using the current fit and the lower age limit, the contrast would be $\sim$12~magnitudes.

\subsection{HD91204}

\begin{table}
\caption{HD91204 Radial-velocity data} \label{tab:rv3}
\center
\begin{tabular}{ccc}
\hline
\multicolumn{1}{c}{JD}& \multicolumn{1}{c}{RV~(m/s)}& \multicolumn{1}{c}{$\sigma$$_{\rm{rv}}$~(m/s)} \\ \hline
& & \\

2451552.102 &   172.2 &  1.5  \\
2451581.998 &   152.6 &  1.4 \\
2451706.816 &   112.0 &  1.5 \\
2451898.177 &    38.8 &  1.5 \\
2451901.157 &    29.2 &  1.5 \\
2451972.066 &     0.0 &  1.4 \\
2451973.023 &     5.4 &  1.4 \\
2451981.986 &    -4.8 &  1.3 \\
2451982.973 &    -1.0 &  1.3 \\
2451983.978 &    -4.3 &  1.4 \\
2452307.969 &  -137.3 &  1.6 \\
2452601.140 &  -254.2 &  1.5 \\
2453017.162 &  -411.7 &  1.5 \\
2453397.979 &  -570.6 &  1.4 \\

\hline
\end{tabular}
\medskip
\end{table}

\begin{figure}
\vspace{2.5cm}
\hspace{-4.0cm}
\includegraphics{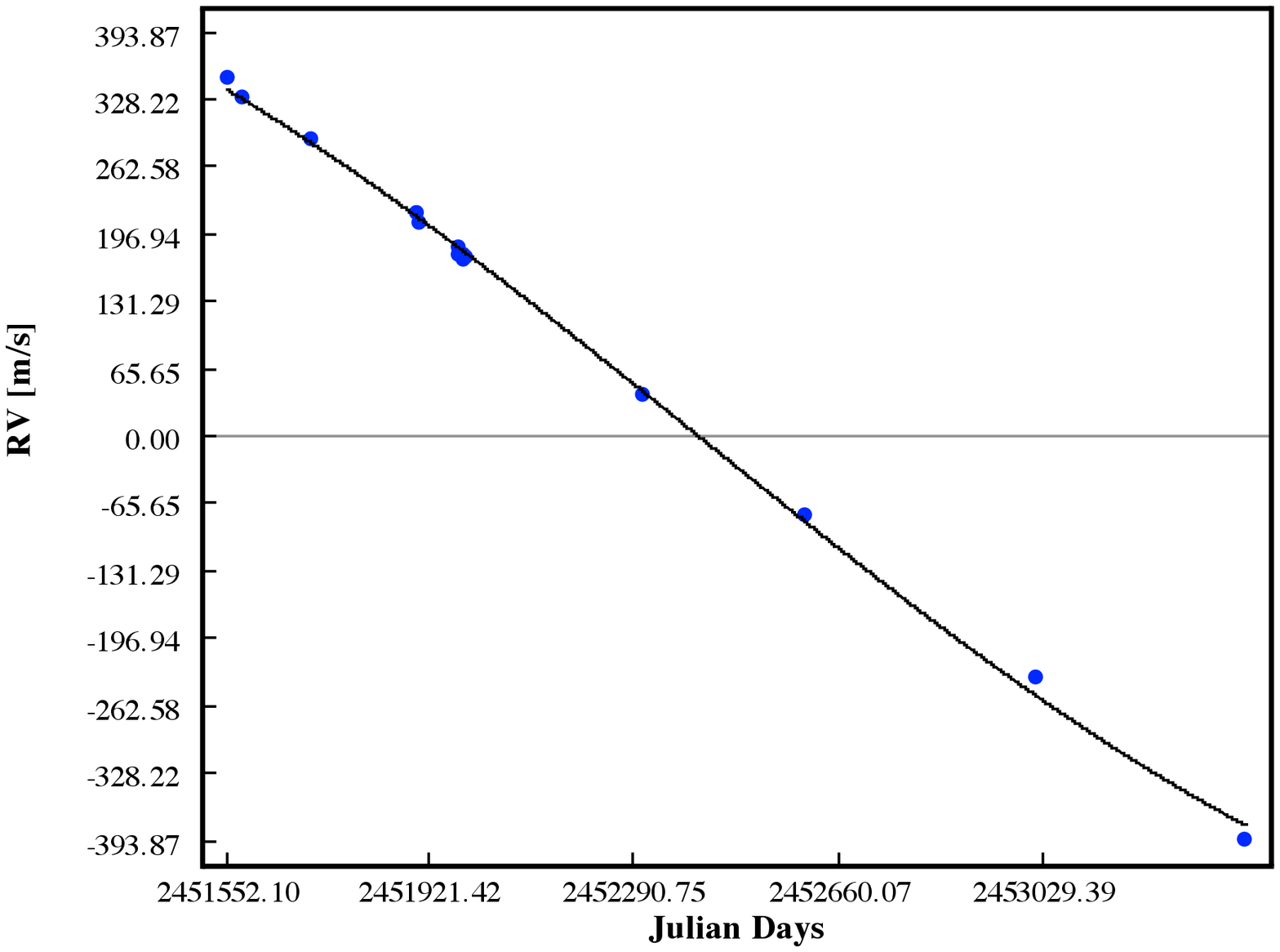}
\vspace{2.2cm}
\caption[RV measurements for HD91204]{The orbital solutions for the radial-velocity dataset from the Keck Planet Search for HD91204.  All 10 data points are spread 
across a period of 5~years and the best
estimate of the fit has an amplitude of $\sim$1000ms$^{-1}$ and a period of $\sim$18~years.  This gives a minimum mass for the companion of $\sim$50M$_{\rm{J}}$.
}
\label{rv_91204}
\end{figure}

Figure~\ref{rv_91204} shows the orbital fit to the large trend in the radial-velocity dataset and again it appears to be a liner.  The estimate to
this data has an amplitude of 1000ms$^{-1}$, a period of $\sim$18~years and circular eccentricity.  The minimum mass of the companion to this fit is $\sim$50M$_{\rm{J}}$ but yet again 
this is a liner fit to the data and therefore the expected orbital period will be underestimated.  Again, twice the orbital period would give rise to a companion with a minimum mass of 
110M$_{\rm{J}}$ but with a fairly high $\chi$$^{2}$ of 15.  From the estimated orbital period above of $\sim$18~years,
and the distance to the star of 51.55pc, we expect the lower limit on the separation to be $\sim$0.13$''$.  The absolute H-band magnitude for HD91204
is 2.83magnitudes with the estimated absolute magnitude for the companion of $\sim$16.05mag, relating to an upper limit for the contrast ratio of 13.22~magnitudes at the given age of 
the system.

\subsection{HD120780}

\begin{table}
\caption{HD120780 Radial-velocity data} \label{tab:rv4}
\center
\begin{tabular}{ccc}
\hline
\multicolumn{1}{c}{JD}& \multicolumn{1}{c}{RV~(m/s)}& \multicolumn{1}{c}{$\sigma$$_{\rm{rv}}$~(m/s)} \\ \hline
& & \\

2452389.145 &    85.9 &  1.4  \\
2452390.076 &    79.7 &  1.3 \\
2452422.026 &    42.3 &  1.4 \\
2452452.991 &     3.8 &  1.1 \\
2452454.920 &    -0.3 &  1.0 \\
2452455.936 &    -0.1 &  1.1 \\
2452509.881 &   -59.6 &  1.7 \\
2452655.135 &  -230.7 &  2.8 \\
2452748.038 &  -345.9 &  1.4 \\
2453217.880 &  -923.7 &  1.9 \\
2453489.100 & -1268.8 &  1.5 \\

\hline
\end{tabular}
\medskip
\end{table}

\begin{figure}
\vspace{3.5cm}
\hspace{-4.0cm}
\includegraphics{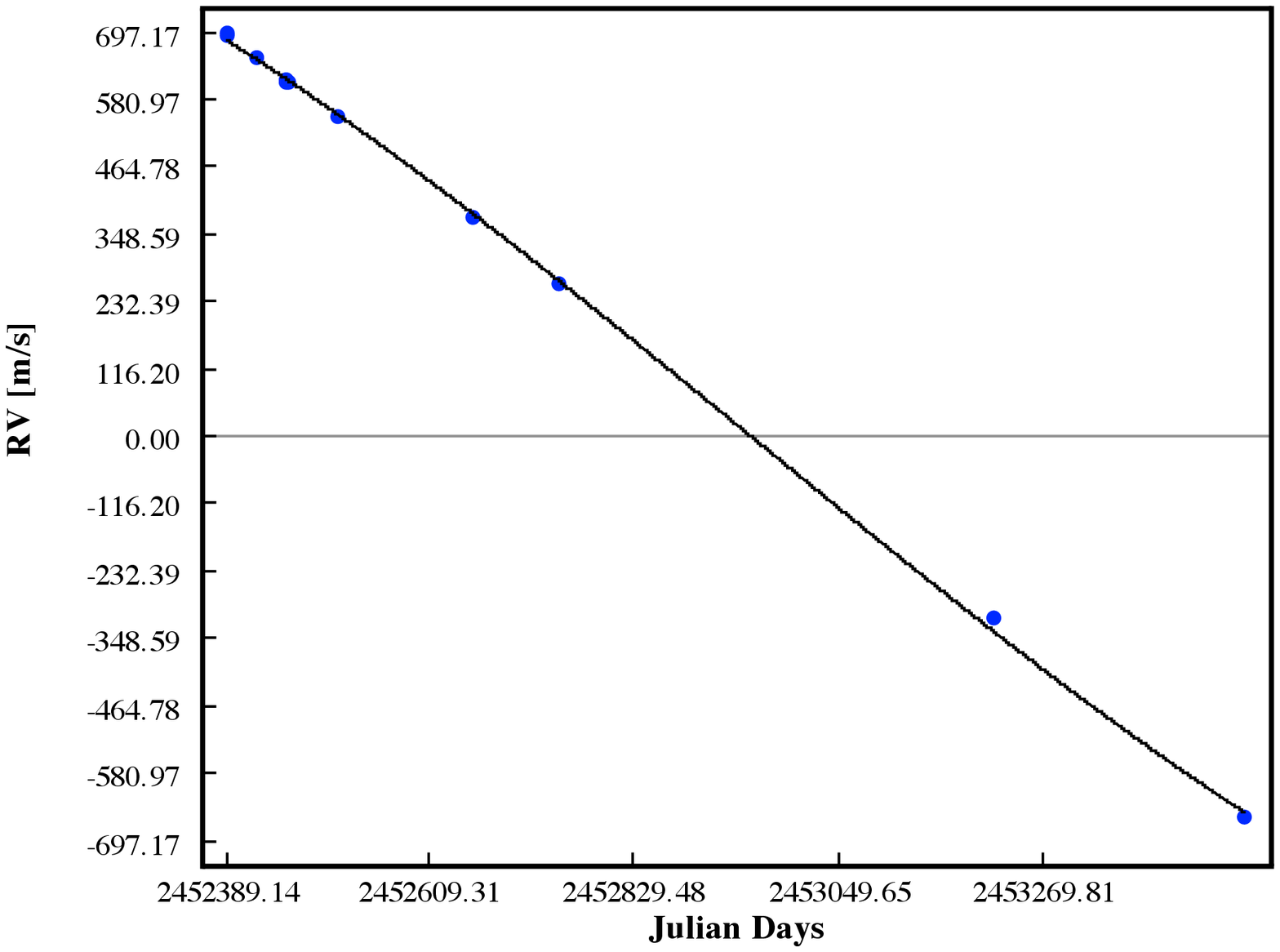}
\vspace{0.7cm}
\caption[RV measurements for HD120780]{The radial-velocity dataset and fits for the companion to the star HD120780.  The minimum-period solution to the large trend is a companion
with a period of $\sim$12~years and minimum mass of $\sim$70M$_{\rm{J}}$, placing it towards the upper end of the brown dwarf regime.  However, this is a liner and the actual 
mass of the companion is probably significantly higher than this.}
\label{rv_120780}
\end{figure}

Eight data points have been acquired over a period of $\sim$3~years and a linear fit has been plotted through the data (Fig.~\ref{rv_120780}).  The best minimum estimate to
the orbital solution gives an orbital period for the companion of $\sim$12~years and assuming a circular orbit, a minimum mass of $\sim$70M$_{\rm{J}}$.  In this case the comparison 
minimum mass for twice the orbital period is 145M$_{\rm{J}}$, with a $\chi$$^{2}$ of 18, much lower than the best minimum estimate.  In fact, searching the parameter space freely for 
the best single companion solution to this data returns a companion with a minimum mass of 1.1M$_{\odot}$ and orbital period of over 56 years.  Clearly such a companion would 
manifest itself in the stellar spectra and this is not the case, which could indicate the need for a double companion solution for this star.  However, taking the 12~year period
we get a semimajor axis of $\sim$5.24AU and at a distance of 17.01pc, the angular separation would be $\sim$0.31$''$.  The absolute H-band magnitude of the star is 4.3magnitudes and
with an estimated lower limit to the absolute H for the companion of $\sim$9.88 magnitudes, the estimated upper limit to the contrast ratio is 5.58 magnitudes.  This represents 
the lowest contrast estimate for the five objects and arises due to the extremely small lower age estimate of 0.2Gyrs from \citet{valenti05}.  Note however that this age is unconstrained 
as the upper age estimate reaches as high as the age of the universe i.e. 13.4Gyrs.

\subsection{HD145825}

\begin{table}
\caption{HD145825 Radial-velocity data} \label{tab:rv5}
\center
\begin{tabular}{ccc}
\hline
\multicolumn{1}{c}{JD}& \multicolumn{1}{c}{RV~(m/s)}& \multicolumn{1}{c}{$\sigma$$_{\rm{rv}}$~(m/s)} \\ \hline
& & \\

2450915.182 &    -222.3  &    2.2 \\
2451002.046  &   -322.3   &   3.0 \\
2451382.974   &  -408.3    &  1.9 \\
2451630.280    & -250.1     & 2.0 \\
2451683.047    & -204.4     & 2.1 \\
2451718.096    & -185.1     & 2.2 \\
2451742.997    & -162.1     & 2.1 \\
2451766.897    & -143.4     & 1.8 \\
2451984.224     & 46.2       &2.3 \\
2452060.982     & 125.7     & 2.0 \\
2452091.945     & 160.8     & 2.1 \\
2452126.927     & 182.4     & 2.4 \\
2452711.315	 &673.1	    &3.1 \\
2452748.215	 &690.2	    &3.5 \\

\hline
\end{tabular}
\medskip
\end{table}

\begin{figure}
\vspace{3.0cm}
\hspace{-4.0cm}
\includegraphics{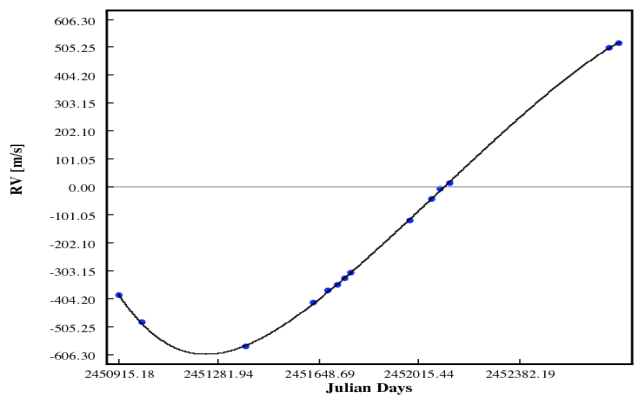}
\vspace{2cm}
\caption[RV measurements of HD145825]{The radial-velocity dataset from the AAPS for the star HD145825.  The best-fit Keplerian orbit to the 14 data points are shown, along 
with the estimated orbital
solutions.  Curvature can clearly be seen in this data, and since this is not a liner the orbital solution is better constrained than all the other four stars.  The
measured semi-amplitude of 617ms$^{-1}$ indicates the presence of a companion with a period of 7.8~years, and eccentricity of 0.2 and a minimum mass of 44.5M$_{\rm{J}}$.}
\label{rv_145825}
\end{figure}

Figure~\ref{rv_145825} shows the 14 radial-velocity measurements made by the AAPS over a period of $\sim$5~years.  This allows a constrained fit to the points as the fitting 
algorithm has one inflection and significant secondary inflection curvature to describe.  The best-fit Keplerian orbit finds solutions that best reproduce the observed curve, with a 
semi-amplitude of 617ms$^{-1}$, described by a companion with 
an orbital period of 7.8~years, an eccentricity of 0.2 and a minimum mass of 44.5M$_{\rm{J}}$.  Since a significant percentage of the orbit has been observed no comparison fit is necessary 
as this fit describes the data very well.  The semimajor axis of this orbit is 4.07AU and at a distance
of 21.55pc the apparent separation of the companion would be $\sim$0.19$''$.  However, even though the solutions are better constrained than the 
liner plots, they are still lower limits as only one inflection is securely found and it is likely that, at best, only 2/3$^{\rm{rd}}$'s of the orbit has been mapped.  The star's absolute H 
magnitude is 3.39~magnitudes and with the best estimates of the companion's absolute H set at 14.2mag, the upper limit on the contrast ratio is estimated to be 10.8~magnitudes.

\section{NACO Imaging}

\subsection{Observations and Reduction}

The observations of each of the stars chosen as primary
candidates were carried out on 02 March 2006 using the
NACO-SDI instrument mounted on the 8m ESO VLT4-Yepun telescope in
Paranal, Chile.
The average seeing throughout the observing night was $\sim$0.8$^{''}$.
The NACO AO system is described in detail in \citet{rousset03}.  
Since all targets in this project are very bright ($V$~$<$~8) the
star itself was chosen as the guide.

The SDI system employs a double calcite Wollaston prism to split the
incoming beam into four separate beams and then feeds them through a quad CH4 filter
that is set in the focal plane.  The filters are set at central wavelengths of
1.575$\mu$m (F1), 1.600$\mu$m (F2) and 1.625$\mu$m (F3a and F3b) and with bandpasses of 0.025$\mu$m, which helps to limit residuals due to speckles and calcite
chromatics.  The \emph{differential} non-common path errors are less than 10nm RMS per Zernicke mode between the beams (\citealp{lenzen04}).
In this configuration the telescope has a projected
field of view of 5~arcseconds square, reduced to 2.7x3.7 arcseconds after the
tilt of 133$^{\circ}$ from the SDI focal plane mask during this run is considered, and has a camera plate scale
of 0.017$''$pixel$^{-1}$. 

The observations were performed using an 8-point jitter pattern.  One of the
jitter frames was a pure sky-frame to better aid in sky background removal.  Each
jitter integration (Detector Integration Time aka DIT) ranged from 2-6
seconds depending on the brightness of the central PSF.  Each DIT was
determined by increasing the time until the central few pixels of
the star were saturated, allowing very high S/N in the halo
of the PSF.  However, we believe that we could further increase this S/N by increasing the DIT's and saturating more of the stellar PSF, since we will gain a higher dynamic range.  Each 
observation took around 60~minutes to complete, consisting of 44 jitter cycles per star.

The reduction of all the NACO-SDI raw data made use of the custom
pipeline of \citet{biller05,biller07}.  First, all the raw frames were cleaned
for any background sky noise by subtracting out the sky image from the
jitter cycle.  A standard flat-field is then applied by combining the
flat images into a master flat and dividing out the pixel-to-pixel variations
from each individual image frame.  To further clean the image a bad pixel map
is created from the jittered images to flag any dead pixels on the CCD chip and
these are removed from all image frames.  Apertures are then extracted
around each filtered image and the Airy pattern and flux is scaled.  The frames
are then unsharp masked by dividing through by a heavily smoothed version of the
original image.  A shift and subtract algorithm is used to align the jittered images, with
the first frame in the series used as the reference image and all other
images aligned to this first image.

\section{Data Analysis}

Once all data frames have been fully reduced the next step is to setup the analysis procedure.  This was done by adding and subtracting the various narrow-band
filters across the methane feature to provide the best conditions for detecting the faint companions.  The two combinations that provide the best contrasts and S/N
ratios to detect cool dwarfs target both M,L-type and T,Y-type objects.  The SDI instrument, by its
pure design, is built to search for companions of mid-T spectral type or later, since the subtraction across the methane
band suppresses the starlight and speckle pattern to highlight objects with strong methane absorption in their atmospheres.  However it can also be used to search for
L-dwarfs by combining all the filters to create a broadband image that would detect any L-dwarf signature.  This, however, is heavily limited by the bright star and the bright
super-speckles in the image.  We employed both these approaches to search for faint companions around the five stars in this project.

\begin{figure}
\vspace{4.5cm}
\includegraphics{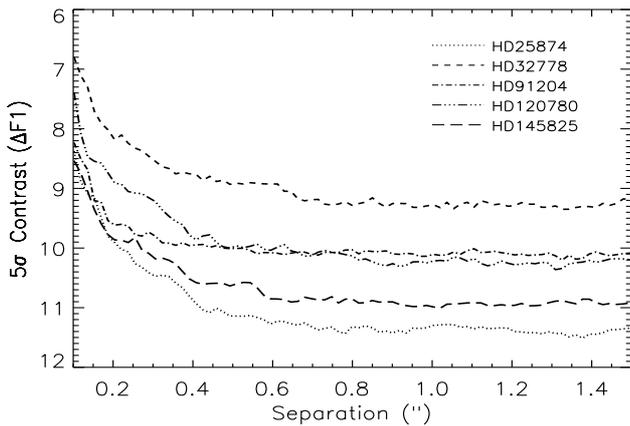}
\vspace{1.35cm}
\caption[Contrast ratio as a function of separation for the cool dwarf candidates]{A combined plot showing all the contrast curves ($\Delta$F1) for each star plotted on a similar 
scale.  The stars and their associated curves are indicated.  It is clear that there is a large spread in contrasts, with both HD25874 and HD145825 
exhibiting similar contrast depths which are $>$2~magnitudes deeper than the shallowest curve, HD32778.  Note however that both HD32778 and HD120780 had their contrasts estimated from 
the other stellar data due to their saturated acquisition frames.}
\label{all_conts}
\end{figure}

Before discussing each system individually, Fig.~\ref{all_conts} shows the SDI reduced contrasts for all the systems on the same plot and scale.  The key in the upper 
right of the plot indicates which curve represents which system.  Since the SDI reduced curve represents the limiting 5$\sigma$ contrast for each system this plot highlights how deep 
the observations reach over the parameter space sampled.  The contrast curves were estimated by defining a 5x5 square pixel box, placing the box at the center of the images and then 
calculating the standard deviation within the box whilst moving it outwards from the center, pixel-by-pixel, in 8 different directions, separated by 45$^{\circ}$ angles.  Averaging 
these gives a measure of the sky background noise.  To measure the contrast ratio, these counts were divided by the peak flux estimate for each star.  Peak flux was 
estimated from the unsaturated acquisition image, scaling to the appropriate exposure time for the saturated data images.  For two of these curves however, representing the stars 
HD32778 (short dashed line) and HD120780 (three dot-single dashed line), there 
are no unsaturated acquisition images and hence their flux counts were estimated using the data from the other stars with unsaturated images.  The 2MASS \emph{H} magnitude for each was 
plotted against peak flux in the acquisition images.  A polynomial was then fit to this data and by taking the \emph{H} magnitude for the saturated stars we could calculate 
their expected peak counts.  This returns an estimate and was only employed since the observational methodology was 
similar.  There is a large spread in contrast limit in this data with a difference of 
$\le$2.2~magnitudes between HD32778 and HD25874 (dotted line), with HD32778 being the least sensitive (max contrast of 9.3~mags) and HD25874 the most sensitive systems (max contrast 
of 11.5~mags).  For all systems the curves are flat beyond 0.7$''$ and remain so out to the edges of the scale.  No additional companions with larger separations than those indicated in 
the radial-velocity datasets were found in any of these systems.  

\subsection{M,L-Dwarfs}

\begin{figure}
\vspace{4.5cm}
\hspace{-4.0cm}
\includegraphics{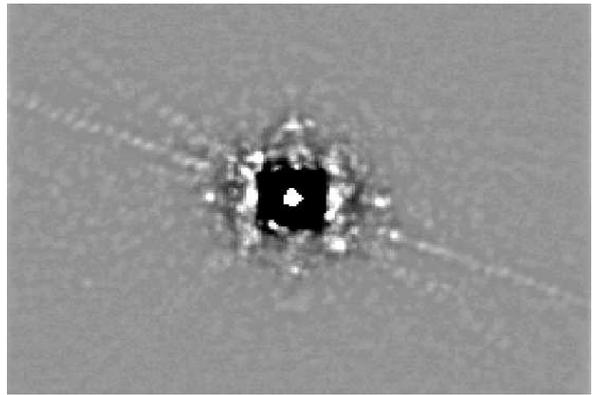}
\vspace{1cm}
\caption[Broadband NACO image of HD25874]{A broadband image of the star HD25874.  The image was constructed by combining all images through the three separate filter bandpasses.  Here 
a number of
super-speckles can clearly be seen scattered across the image, the brightest of them reside close to the central star.  All the possible cool dwarf companions visible in this image are ruled out as speckles when comparing the rotated images.}
\label{broadband}
\end{figure}

As mentioned above to extract the signal from companions that don't exhibit strong methane in their atmospheres the best SDI approach is to combine all the narrow-band
filters employed by the SDI device and create a broadband image that will increase the S/N of any cool companion.  This method was applied to the data
frames from all five of the target stars.  Note that a targeted search for such objects that don't exhibit strong methane absorption in their atmospheres would benefit significantly 
from utilising a standard broadband AO approach, since with the narrowband SDI filters there is a loss of efficiency and hence a loss of contrast for a given total integration time.  
To help with the detection of any signals and to decipher them from residual stellar speckles all images were taken at two
different rotation angles, separated by 33$^{\circ}$ on the sky.  By blinking these rotated broadband images it is possible to spot any detection that appears at
different positions in the rotated images.  Any super-speckles should remain in the same place throughout the roll angle, therefore a real detection can be picked
out in a field with a number of residual speckles.  Figure~\ref{broadband} is an example of one of these broadband images for the star HD25874.

From the estimates of the companion minimum masses generated from the orbital fits we might expect three of these companions to reside in the L-dwarf regime and considering the limited 
data they may also have M dwarf masses.  However
the estimates of the contrast ratios are high and as the contrast that can be reached by using a broadband image is significantly lower than the subtracted images,
it would prove extremely difficult to detect any of these L-dwarfs using this method.  Indeed, the search employed for L-dwarfs around all five of these stars turned up
a number of bright objects, these can be seen as the spots spread across the image in Fig.~\ref{broadband}.  However, none of these detections fulfilled the rotation roll
angle criteria, indicating they are residual speckles.  It is clear that without looking for roll angle modulation it is extremely difficult to differentiate between a speckle and a 
real object.  

\begin{figure}
\vspace{4.5cm}
\hspace{-4.0cm}
\includegraphics{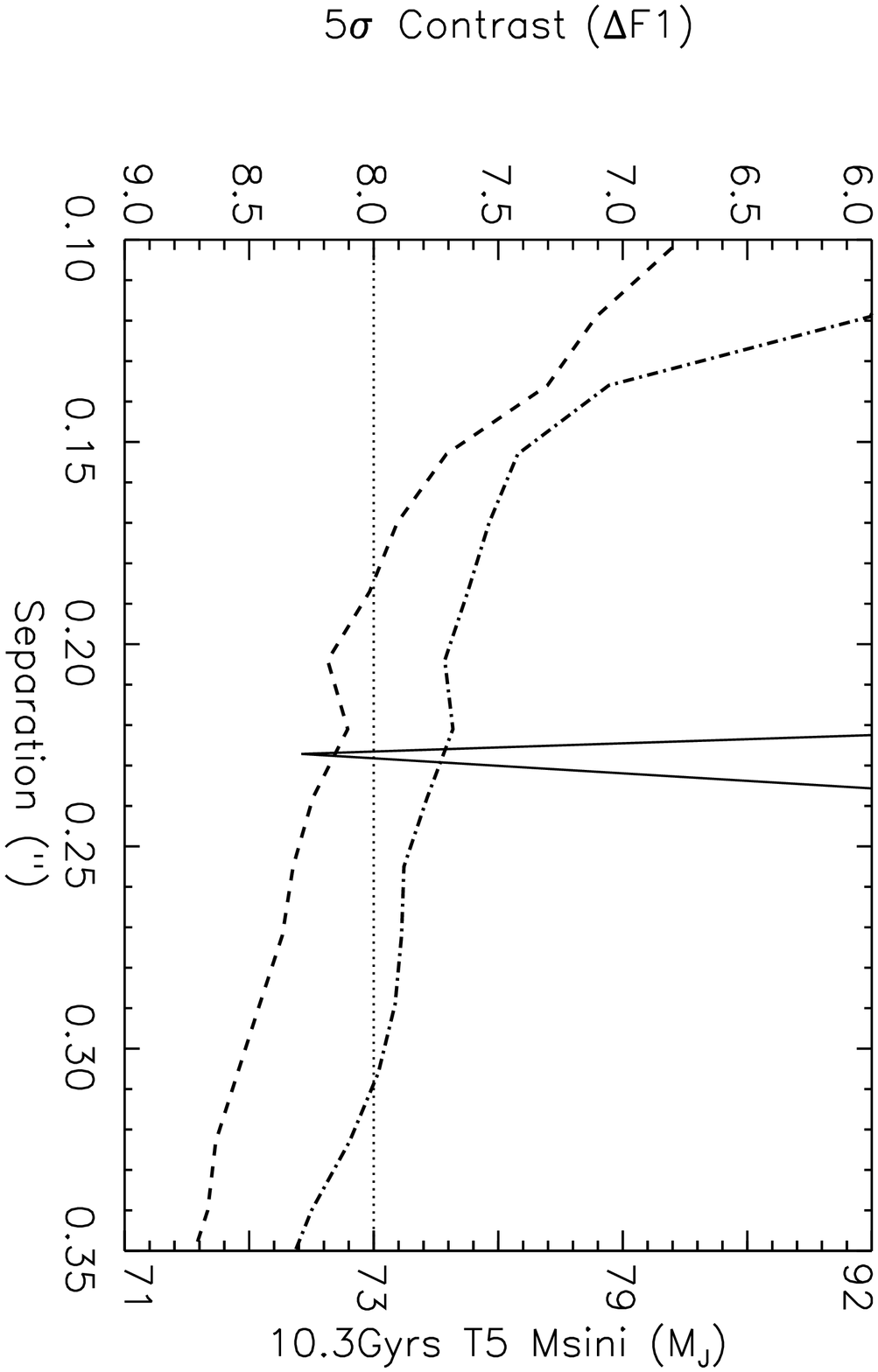}
\vspace{4.5cm}
\includegraphics{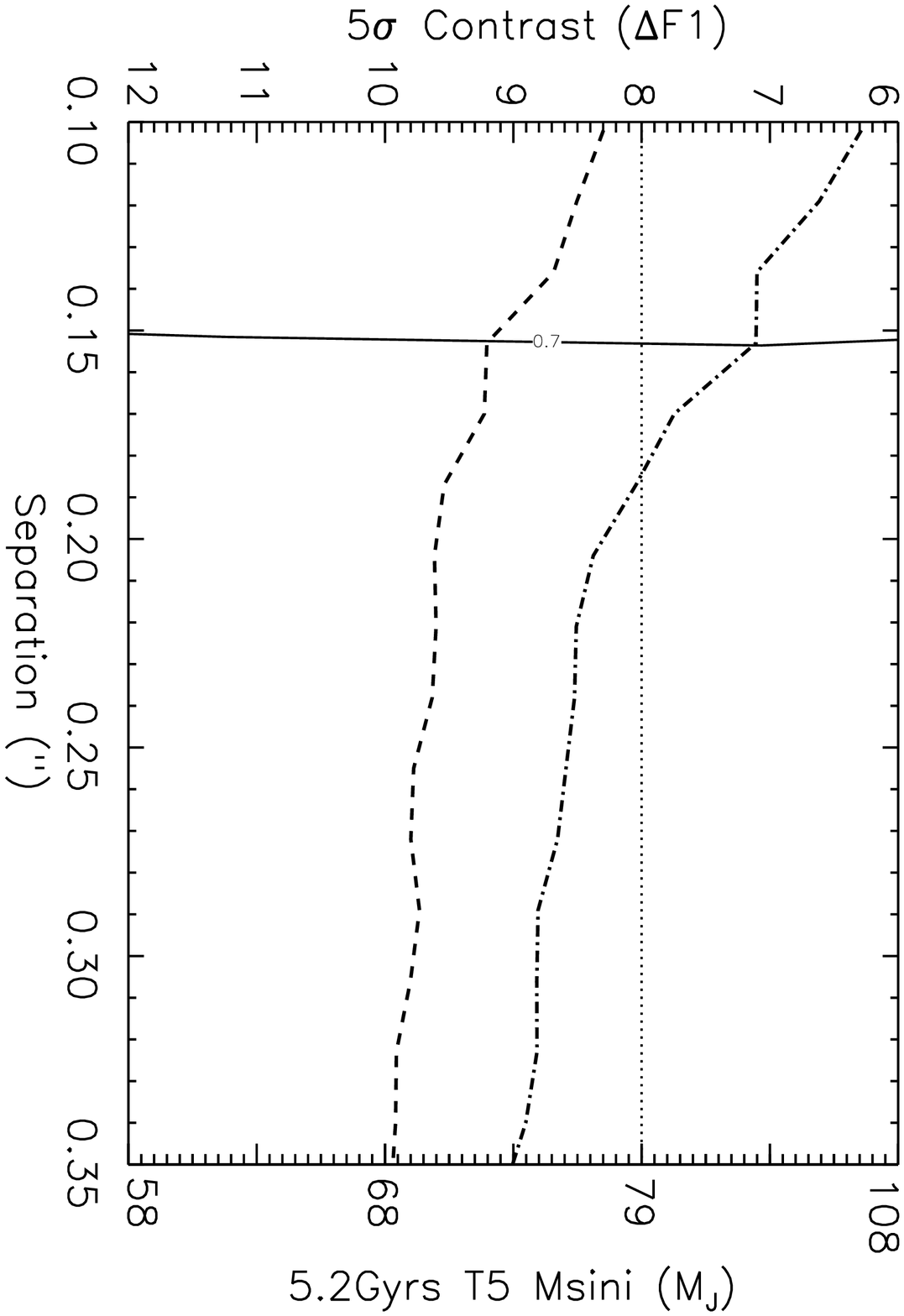}
\vspace{4.5cm}
\includegraphics{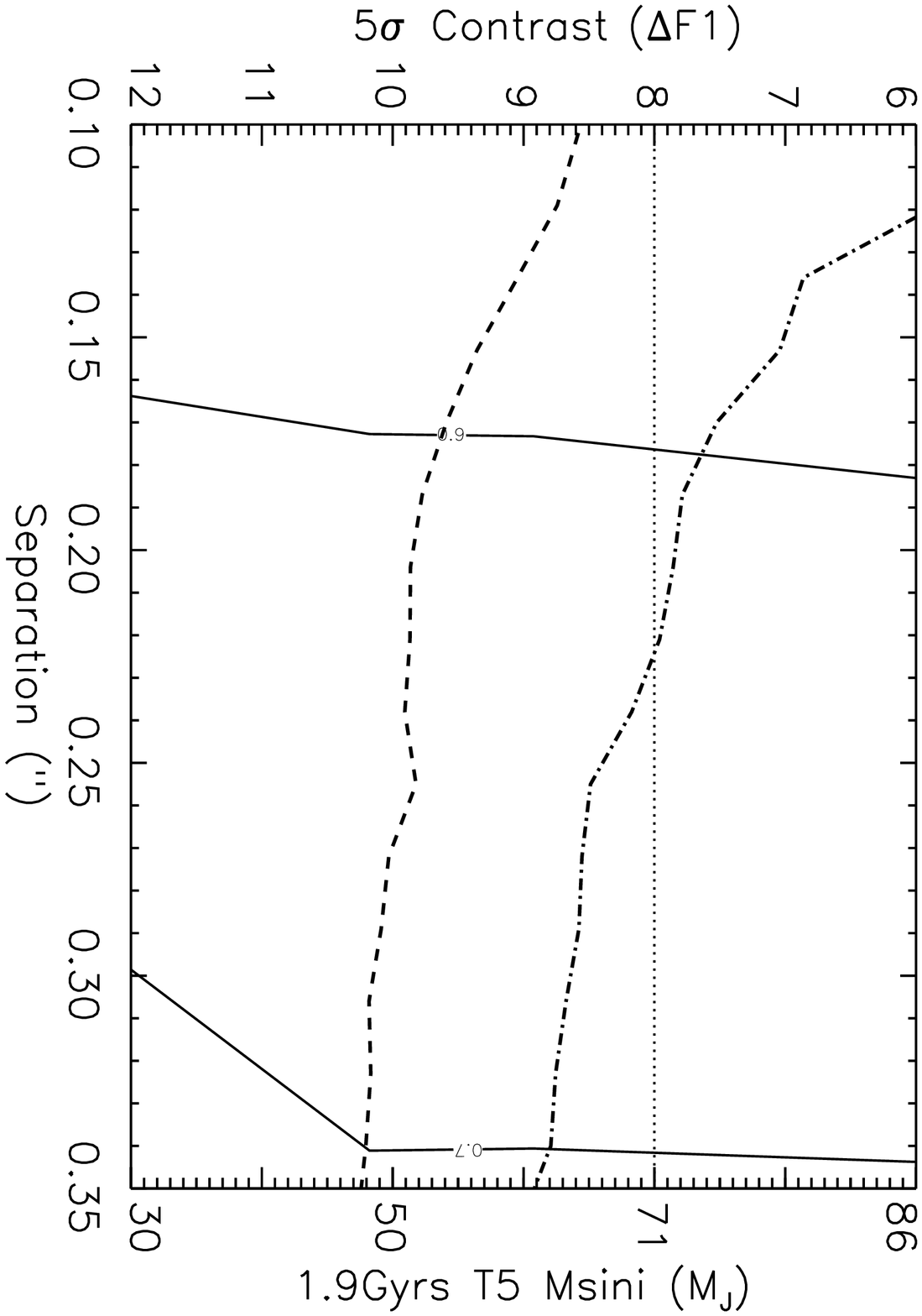}
\vspace{2.8cm}
\caption[Contrast curves for HD32778, HD91204 and HD145825]{The 5$\sigma$ contrast limits ($\Delta$F1) for the stars HD32778 (top panel), HD91204 (middle panel) and 
HD145825 (bottom panel).  The 
dot-dashed curves represent the Conventional AO detection thresholds and the dashed curves represent the SDI reduced thresholds.  No detections 
were found around any of these stars.  The solid curves represent the confidence limits we are sensitive too from the radial-velocity data sets.  Both HD32778 and HD91204 cover only 
a limited parameter space given by the radial-velocity data, 
which means the companion is still relatively unconstrained.  Whereas, HD145825 observations reach deeper and are able to rule out a lot of solutions for the companion.  The right-hand 
axes show companion masses for a T5 dwarf at the best estimated age of the systems from \citet{valenti05} and \citet{takeda07} and are for comparison only.  The 
horizontal dotted line represents the approximate strong methane boundary.}
\label{contrasts}
\end{figure}

If the companions to these stars are L-type objects and the estimated separations from the radial-velocity 
curves are 0.13$''$ (HD25874), 0.16$''$ (HD32778), 0.13$''$ (HD91204), 0.31$''$ (HD120780) and 0.19$''$ (HD145825) then they must not have contrasts less than 
6.7,~7.4,~7.0,~7.9~and~7.8 
magnitudes respectively (taken from the conventional AO (dot-dashed) curves in Figs.~\ref{contrasts},~\ref{hd25874_contrast}~and~\ref{hd120780_contrast}).  Note that 
the right-hand y-axes in 
these figures show the estimated masses at the age of these systems relating to contrast limits for methane objects only and this will be explained in $\S$5.2.  At the estimated 
separations the mass limits for L-dwarfs are 87,~77,~92,~72~and~72 
Jupiter-masses respectively.  No L-type objects with the given ages and separations can have masses above these limits.  Also shown in these plots represented by the solid 
curves are the radial-velocity sensitivities.  These confidence limits from the radial-velocity data sets are explained below and the conclusions drawn are presented there.

\subsection{T or Y-Dwarfs and Exoplanets}

The search for T and Y-dwarfs, along with planetary mass objects, cooler than 1200K is accomplished by subtracting the images inside and outside of the methane band at 1.62$\mu$m.  
The filters employed for this
task are the F1 filter centered at 1.575$\mu$m and the F3a filter centered at 1.625$\mu$m (\citealp{close05}).  This subtraction, which we assign $\Delta$F1, helps to suppress the Airy 
rings from the stellar
PSF, and attenuates the speckle pattern that is a function of the stellar beam.  As there is no strong absorption at 1.62$\mu$m in the stellar atmosphere, the subtraction 
removes the stellar light leaving behind the faint methane signature from
the cool companion.  Similar to the search for M,L-type objects the 33$^{\circ}$ roll angle allows one to distinguish between a genuine cool companion signal and any
residual speckles.

For each of the five stars we performed this search on, two candidates fulfilled the roll angle criteria, found around the stars HD25874 and HD120780, both of which are discussed in the 
next section.  For the three other stars no significant companions are seen in the subtracted images across the roll angle.  The SDI reduced contrast limits in Fig.~\ref{contrasts} 
(dashed line) for the stars HD32778, HD91204 and HD145825 tell us that if the radial-velocity companions that
were targeted were T or Y-dwarfs, they can not have absolute H magnitudes $<\sim$11.1, 11.5 and 13.2 magnitudes respectively, at the estimated separations for these objects.  
All $H$, M$_{\rm{H}}$ and mass estimates are assuming T5-type status, including the right-hand y-axes, which show the expected masses for the corresponding contrast ratios at the 
estimated age of the star.  We chose a T5 simply because it represents 
the middle of the T-dwarf regime and since these objects are expected to be old we might expect them to reside somewhere close to this spectral type.  A reference such as a T dwarf 
was necessary due to the unique nature of the SDI observations, since they utilize narrow-band filters across a strong absorption feature and objects with more or less methane 
absorption with have more or less flux entering the F3a filter.  Since T5 objects possess a strong methane break they provide good references for comparison in our contrast curves.  
A correction must be made to compare with model absolute $H$-band magnitudes and we simply apply an offset 
of +0.5~magnitudes for a typical T5 dwarf (\citealp{biller07}).  

\subsection{Doppler Sensitivities}

\begin{figure}
\vspace{4.5cm}
\hspace{-4.0cm}
\includegraphics{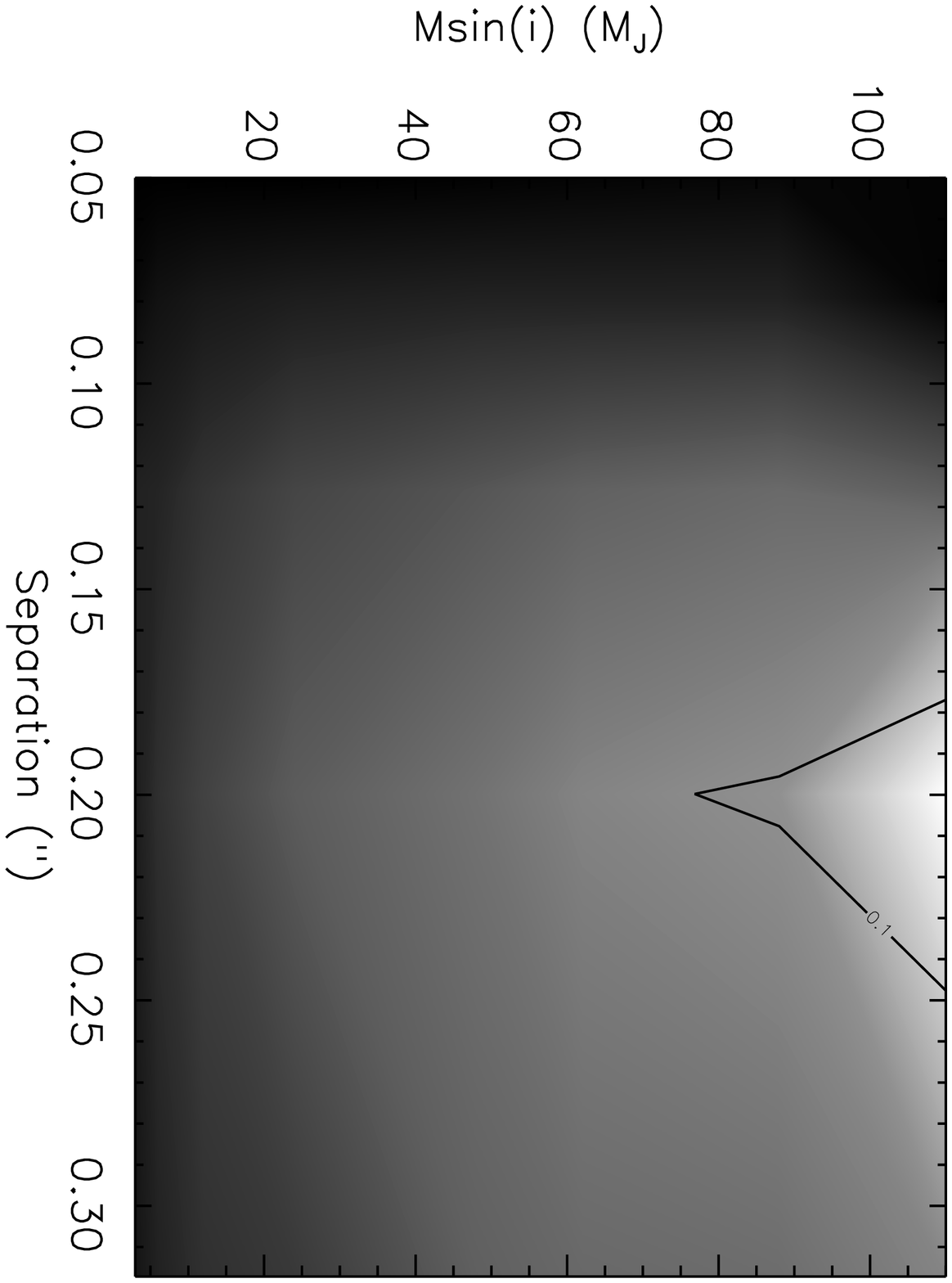}
\vspace{4.5cm}
\includegraphics{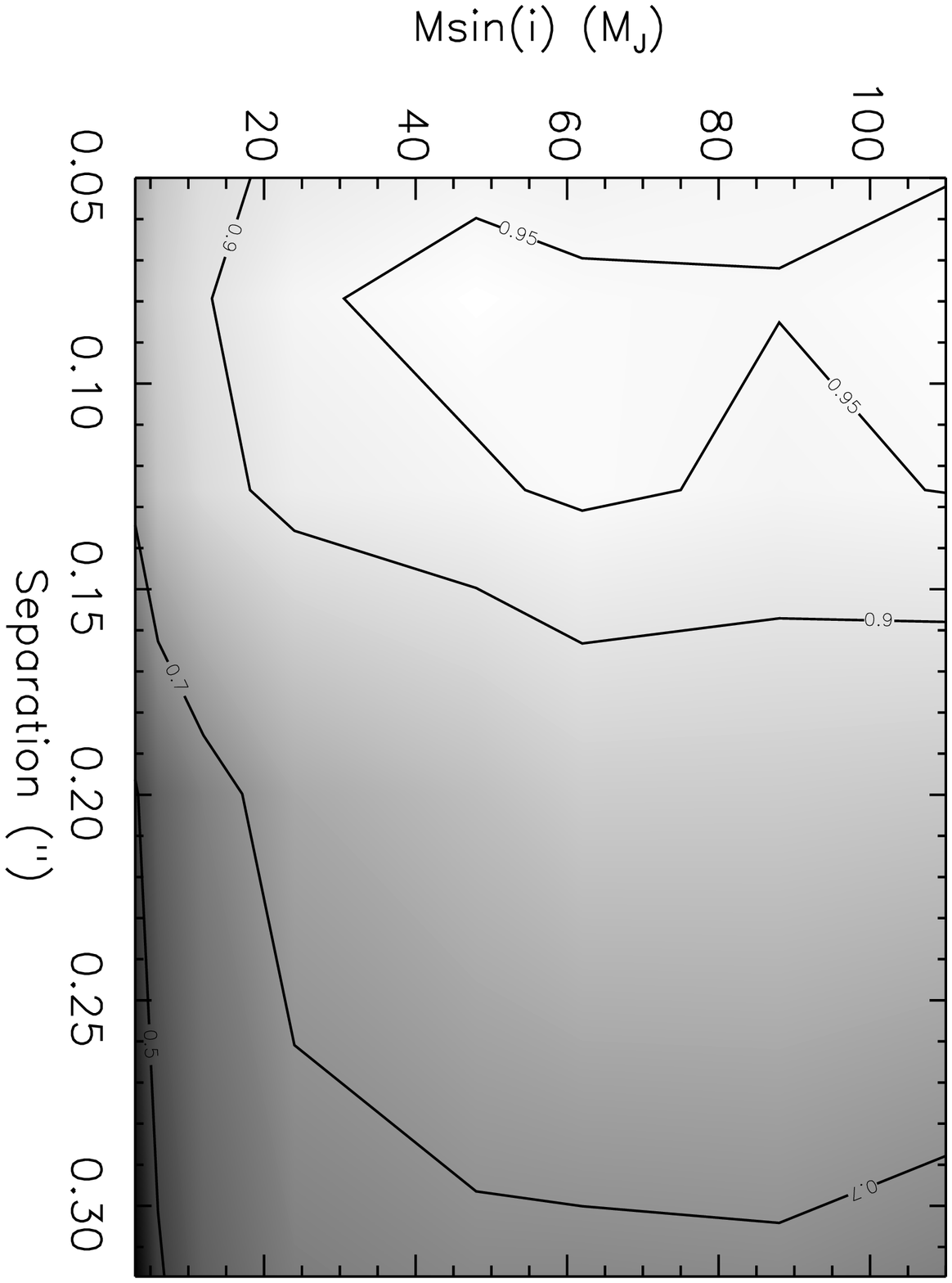}
\vspace{4.5cm}
\includegraphics{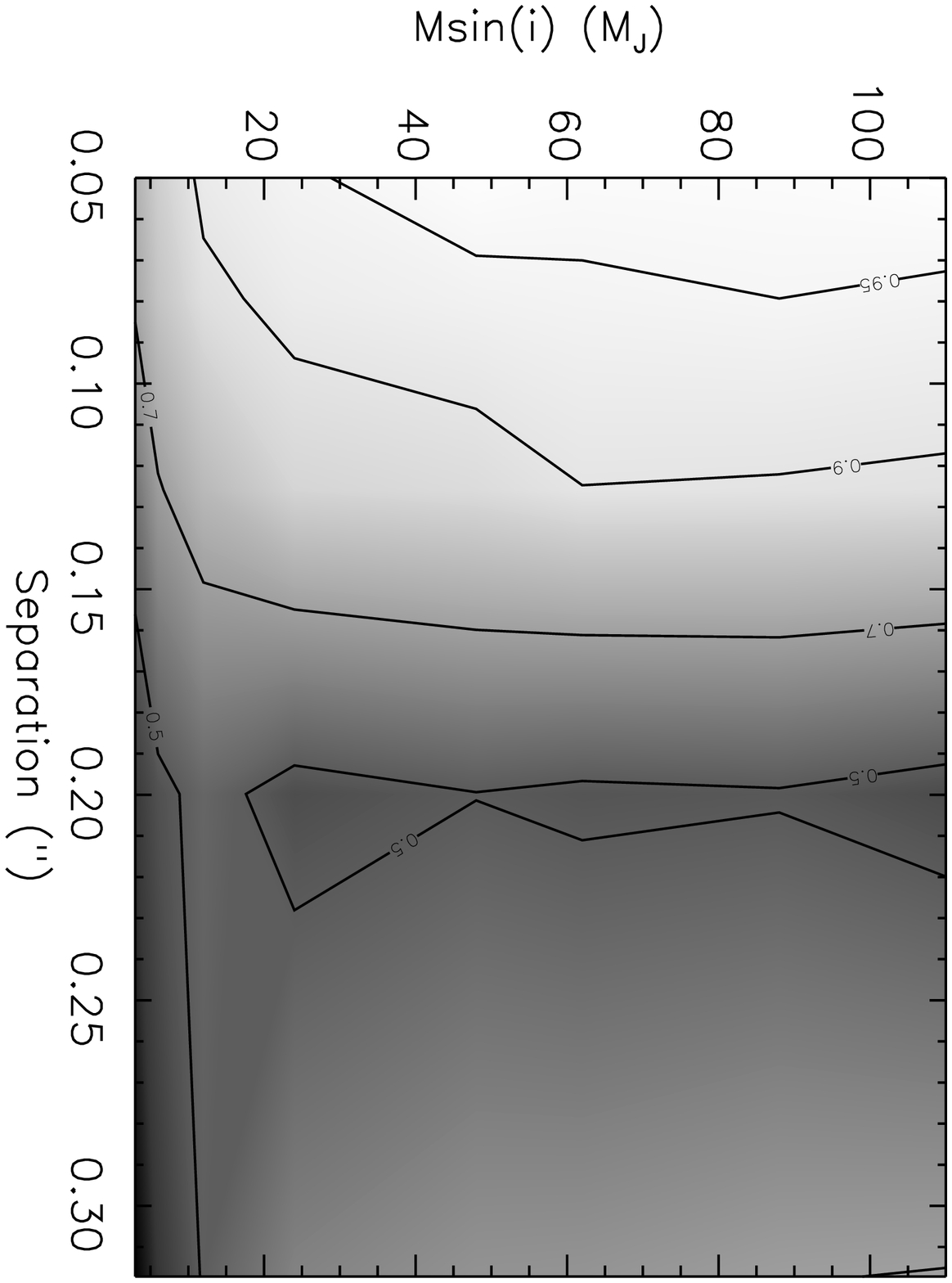}
\vspace{2.8cm}
\caption[Detectabilities for HD32778, HD91204 and HD145825]{The doppler sensitivity confidence limits in mass-angular separation space for the stars HD32778 (top panel), HD91204 
(middle panel) and HD145825 (bottom panel).  These confidence limits have been integrated over all eccentricities and the percentage limit is shown inside each curve.  The gray scale 
highlights the changing spacial confidence limits with dark the least constrained through to the most constrained regions being lightest.  Only 
HD145825 has fairly well constrained doppler data out to reasonable NACO-SDI sensitive separations.  Whereas for HD32778 there is no real constrained parameters below the 
hydrogen burning limit.}
\label{detects}
\end{figure}

We have performed a sensitivity analysis of each individual system to determine the confidence limits from the 
radial-velocity datasets and how these limits transfer into the contrast curve parameter space.  The sensitivities were determined using the method explained in \citet{otoole09} 
with a small modification. In brief, simulations of radial-velocity curves are analysed using a two-dimensional Lomb-Scargle
periodogram. The simulations covered the mass range 3-110M$_{\rm{J}}$, period range 300-4800\,d and eccentricities from
0.0 to 0.8 in steps of 0.1.  Jitter values for each star, based on the method of \citet{wright05}, were incorporated into the simulations.  
The resulting fits were checked against five detection criteria: the four described in O'Toole et al. and a fifth based on the RMS of the
data, described here.  If the measured period is 50\% longer than the time baseline of the dataset, a linear fit to the dataset is performed.  This 
fit is then subtracted from the data and the RMS of the residuals are compared to the RMS before the linear trend was subtracted.  A detection is made if the RMS drops by a factor two or 
more throughout this process.  When used in conjunction with the SDI contrast curves, these doppler sensitivies can help to more accurately constrain the parameter space that we have 
covered and help us to determine what possible companion objects the data did not detect.  Sensitivities are shown as solid curves in Figs.~\ref{contrasts} and \ref{detects}.

The radial-velocity confidence limits for HD32778 (Fig.~\ref{contrasts} top) show that due to the lack of data points for this star we have no strong constaint on any 
sub-stellar companions.  The 
curve here represents the 10\% confidence limit for this star (shown more clearly in the top plot of Fig.~\ref{detects}) and therefore below the hydrogen burning limit we have no 
real constraints on any companions in this parameter space.  The gray scale in this figure, and the following detectability curves, show the changing confidence limits running from dark 
being the least constrained to light the most constrained regions.  Given the lack of data here there is only a very small region where we can rule out any companions with an confidence 
and most of the parameter space is highly unconstrained.  The 10\% confidence bound barely reaches beyond the 5$\sigma$ (99.9\% confidence limit) of the SDI reduced 
contrast curve and only covers 
a very small part of the parameter space beyond the AO curve.  Hence, for this system the radial-velocity data does not help to constrain the companion and any information we 
can extract about the unseen companion is drawn from the contrast curves alone.  This leads to the probability that the companion is of lower mass 
($\sim$$\le$~70M$_{\rm{J}}$) with a fairly unconstrained orbital separation i.e $>$0.076$''$ from the baseline of the radial-velocity data.

Similar to HD32778, the confidence limits in mass-separation space for HD91204 are poorly constrained (Fig.~\ref{contrasts} middle).  However, since HD91204 has a larger database of velocities 
than HD32778, 
we can say to a 70\% confidence level that the companion to this star is not a close by ($\le$0.15$''$) sub-stellar or stellar secondary.  The middle 
panel in Fig.~\ref{detects} shows the mass-separation parameter space for this star, reaching down onto the stars surface.  Clearly very close-by companions can be ruled out to high 
levels of 
confidence due to the larger number of data points and the fact that this radial-velocity curve is a liner.  This is further highlighted by the large light region shown in the gray scale and a 
lack of any large contrast gradient.  Objects with separations below 0.06$''$ and masses above around 30M$_{\rm{J}}$ can be ruled out to $\sim$90-95\% confidence and moving out to 
separations of 0.08$''$ we can still rule out objects down to around the planetary mass limit at the 90\% level of 
confidence.  Therefore, we can say that the companion to this star likely has a fairly large separation and is a faint substellar companion or, as for almost all of the imaged objects, the 
companion has a longer orbital period, but the inclination of the system was such that when we were observing the companion it was hidden behind the star at on-sky angular separations 
below 0.1$''$. 

In comparison to these other two stars, HD145825 has enough data points and exhibits enough curvature in the timeseries that fairly high levels of confidence from the velocities 
overlap with the 99.9\% confidence limits from the imaging work.  The bottom panel in Fig.~\ref{contrasts} shows that below the SDI curve we still have over 90\% confidence in ruling out 
close by ($\le$0.17$''$) objects down to low brown dwarf masses.  In addition, the 95\% (2$\sigma$) confidence limit can rule out a lot of possible brown dwarf/stellar companions below 
the 0.1$''$ angular separation limit of the SDI technique (Fig.~\ref{detects} bottom).  The gray scale reveals more structure than HD32778 and HD91204 due to the significant curvature in 
the velocities.  Particularly we can see that the region around 0.2$''$ separation is less constrained than inside and outside this separation and due to the indication of secondary 
curvature in the velocities, we arrive at fairly high confidence levels beyond 0.3$''$ separation.

From these combined constraints we can rule out to really high levels of confidence any brown dwarf/stellar 
companions with small separations (short period orbits).  Also, at the 1$\sigma$ level we can say that there are no objects at all below a separation of around 0.34$''$ with masses above 
40M$_{\rm{J}}$ and also no companions down into the giant exoplanet regime within 0.20$''$.  These combined data sets argue for the companion 
to HD145825 to be an extremely faint sub-stellar companion with a moderate separation.  

\section{HD25874 \& HD120780 Detections?}

Out of the five stars that we searched around, two possible detections were made around the stars HD25874 and HD120780.  Both of these candidates fulfilled the
requirements to be considered as bona fide candidates as they were bright sources, that had counterparts at 33$^{\circ}$ in the rolled images.  However, after careful analysis we 
believe these to be artifacts of the reduction procedure and not true companion objects.  Both will be discussed here.

\begin{figure}
\vspace{2.5cm}
\hspace{-4.0cm}
\includegraphics{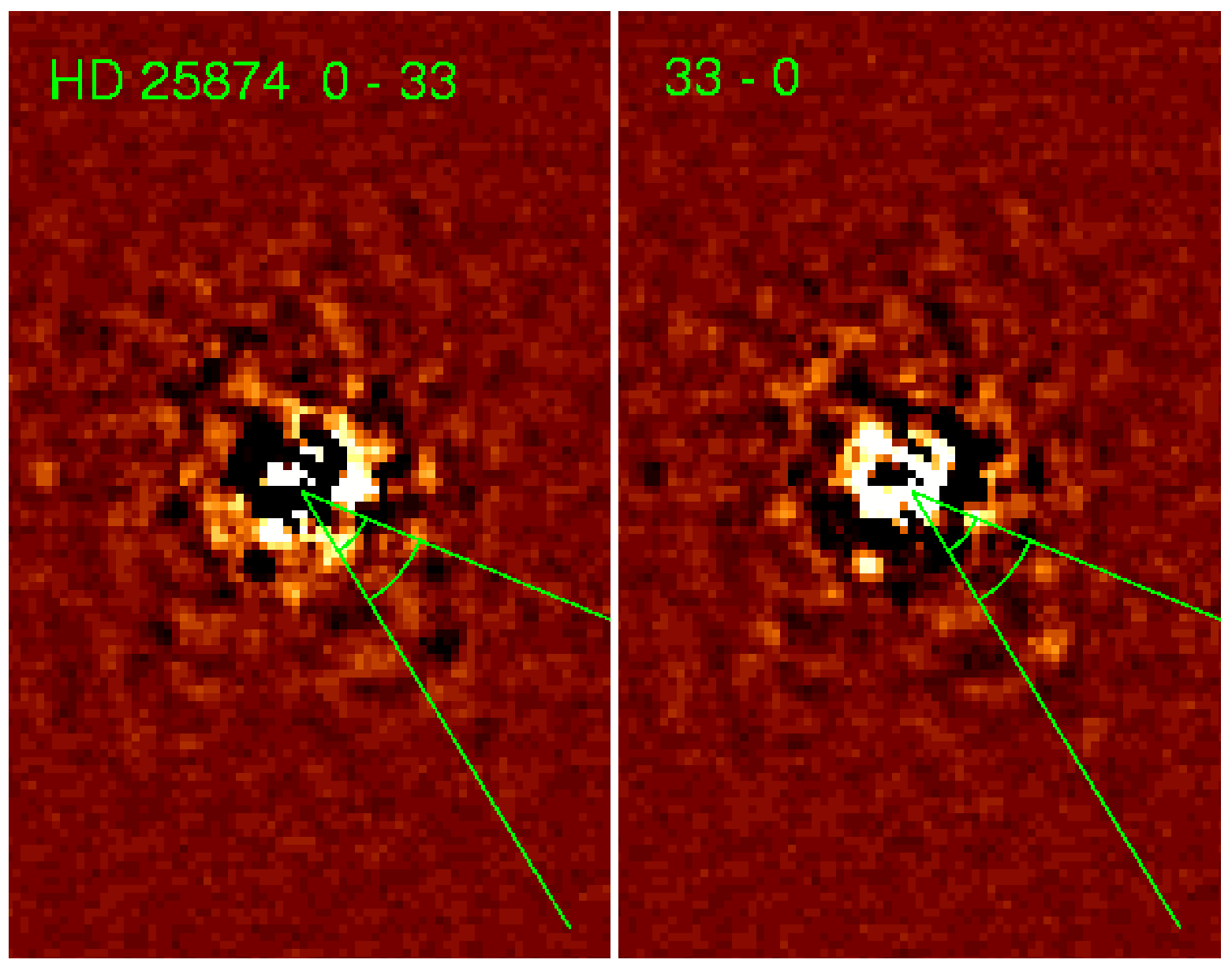}
\vspace{3.3cm}
\caption[Annotated image of HD25874]{An annotated image of the star HD25874 at both inverted roll angles.  The lines mark the 33$^{\circ}$ projection of rotation of the 
camera.  The negative and positive detections of the possible source companion intersects both these lines at the projected separation of $\sim0.29''$.}
\label{annotate_hd25874}
\end{figure}

Figure~\ref{annotate_hd25874} shows the 33$^{\circ}$ roll angle of the camera and how it projects along the image through the T3 filter (F1(1.575$\mu$m)-F3a(1.625$\mu$m)) 
for HD25874.  The detections are found at the ends of the second
arc along the projection with a separation from the central pixel of 0.29$\pm$0.01$''$ and a position angle of 240$^{\circ}$.  This enhanced image highlights more of the 
speckles across the images in both negative and positive formats e.g. bright spot to the extreme
left middle of the left panel and its counterpart in the corresponding position of the right panel.  Along the projected angle there is also another bright and dark pair
that could be separated by the roll angle and these are found at the ends of the inner arc.  As these are so close and connected to the central star we believe these to be 
an artifact of the PSF subtraction, however worryingly since they are found projected along the same axis as the potential candidate detection they may signify that the 
detection is an artifact as well e.g. uncorrected residual trefoil in the image.

\begin{figure}
\vspace{2.5cm}
\hspace{-4.0cm}
\includegraphics{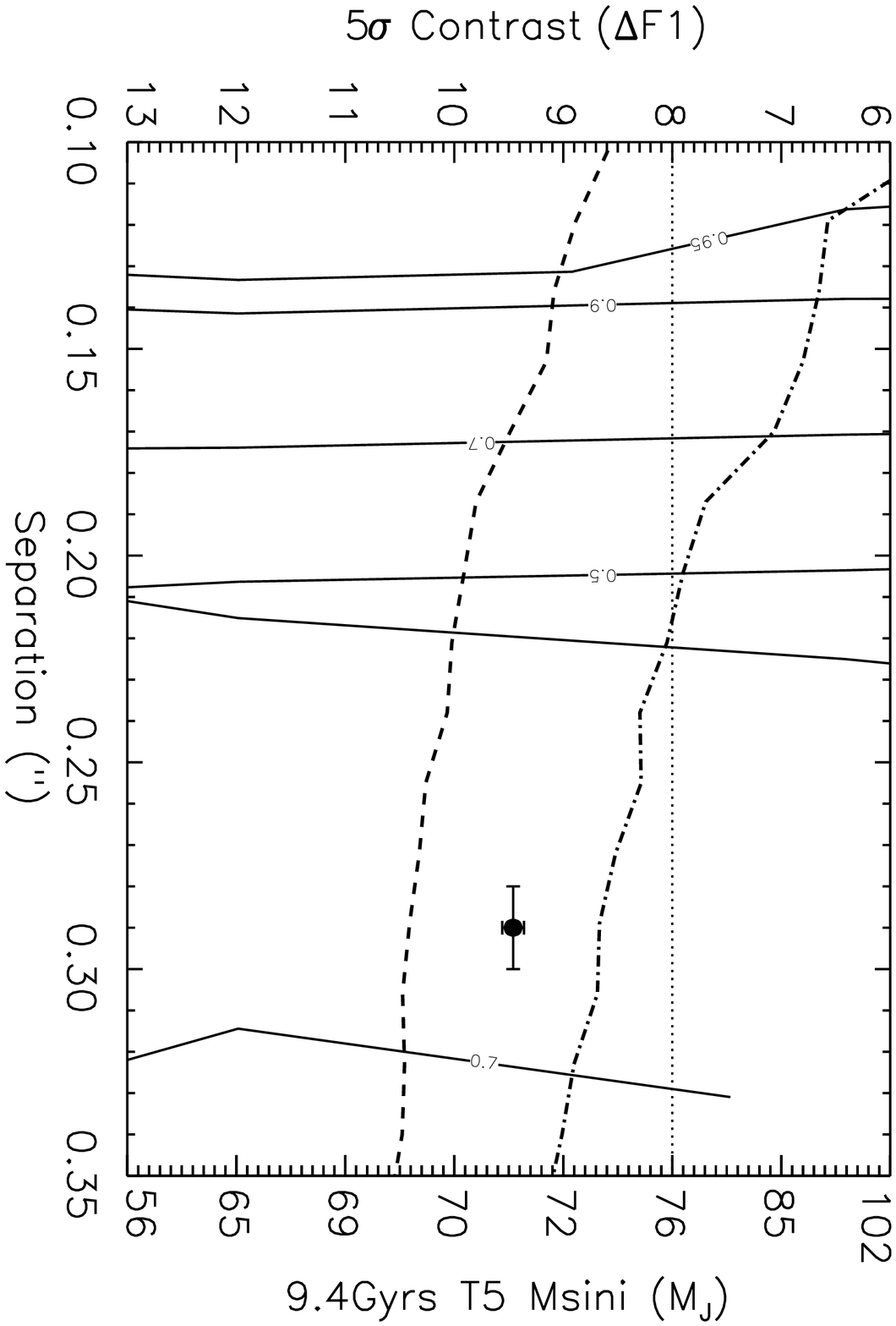}
\vspace{5.5cm}
\includegraphics{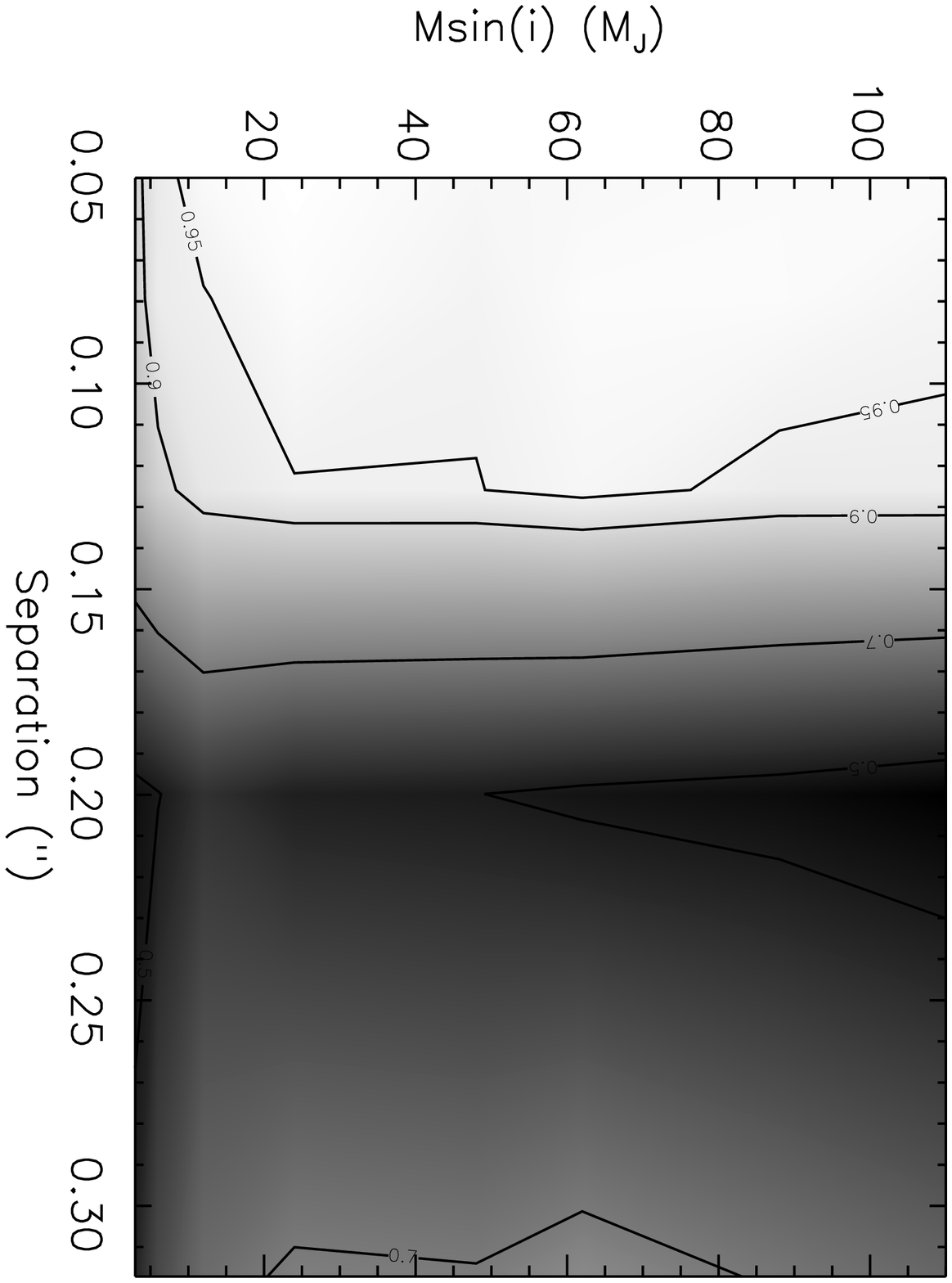}
\vspace{1.4cm}
\caption[Contrast limits for HD25874]{The upper plot shows the contrast limits ($\Delta$F1) for the star HD25874.  A contrast of 10.5~magnitudes is reached at 0.3$''$, which is 
the separation of the potential candidate.  The solid curves mark the radial-velocity confidence limits.  The right hand y-axis shows the expected mass for a T5 dwarf at the age assumed 
for these objects.  The horizontal dotted line marks the strong methane boundary.  The best fit above this limit is to aid the eye since no detections were made beyond this boundary.   The 
lower plot shows the radial-velocity confidence limits in mass-angular separation space.  The values inside all solid curves are the confidence percentages.  }
\label{hd25874_contrast}
\end{figure}

As mentioned, Fig.~\ref{hd25874_contrast} (upper) shows the contrast limits that were determined for HD25874, highlighting both the conventional AO and the SDI reduced limits.  For 
companion candidates such as the one here it is clear that the SDI reduction performs significantly better than conventional AO e.g. gain of $\sim$2.5~magnitudes at 0.2$''$.  The 
confidence limits in this figure show a lot of structure at separations reaching well into the imaging phase space.  This is due to the large baseline ($>$4~yrs) of observations, even 
though they describe a liner system.  95\% confidence limits are seen to rule out objects down into the planetary mass regime with angular separations below 0.14$''$, depending on where 
we place the boundary between exoplanets and brown dwarfs, and we can rule out 
such objects up to separations of almost 0.18$''$ at the 70\% confidence limit.  The lower panel better highlights the structure in detectability for this star and shows 
clearly that we have high levels of confidence out to the timeseries of this data set.  The gray scale shows similar 
structure to that of HD145825 with a large inner region that is highly constrained, a dark unconstrained region around 0.2$''$ separation from the star and then a growing lighter 
constrained region out to separations of 0.3$''$.  This time the constrained region at larger separations arises not from any curvature in the velocities but due to the overall span of 
velocity across the data set, showing that it is unlikely that lower mass companions at these orbital separations could give rise to this data set.  By combining 
the imaging data with the radial-velocity data, we can say that with 70\% confidence we can rule out almost all companions to this star with separations below 0.17$''$ (4.40~AU).  
Again we conclude that the companion to this star is probably a widely separated, low mass ($<$70M$_{\rm{J}}$) and therefore really faint sub-stellar object.  If such is the case 
then there is a fairly high possibility here that the object is a brown dwarf located in the brown dwarf desert (\citealp{grether06}).  Note that there were no other objects detected 
around this star out to orbital distances of $\sim$52~AU.

\begin{figure}
\vspace{2.5cm}
\hspace{-4.0cm}
\includegraphics{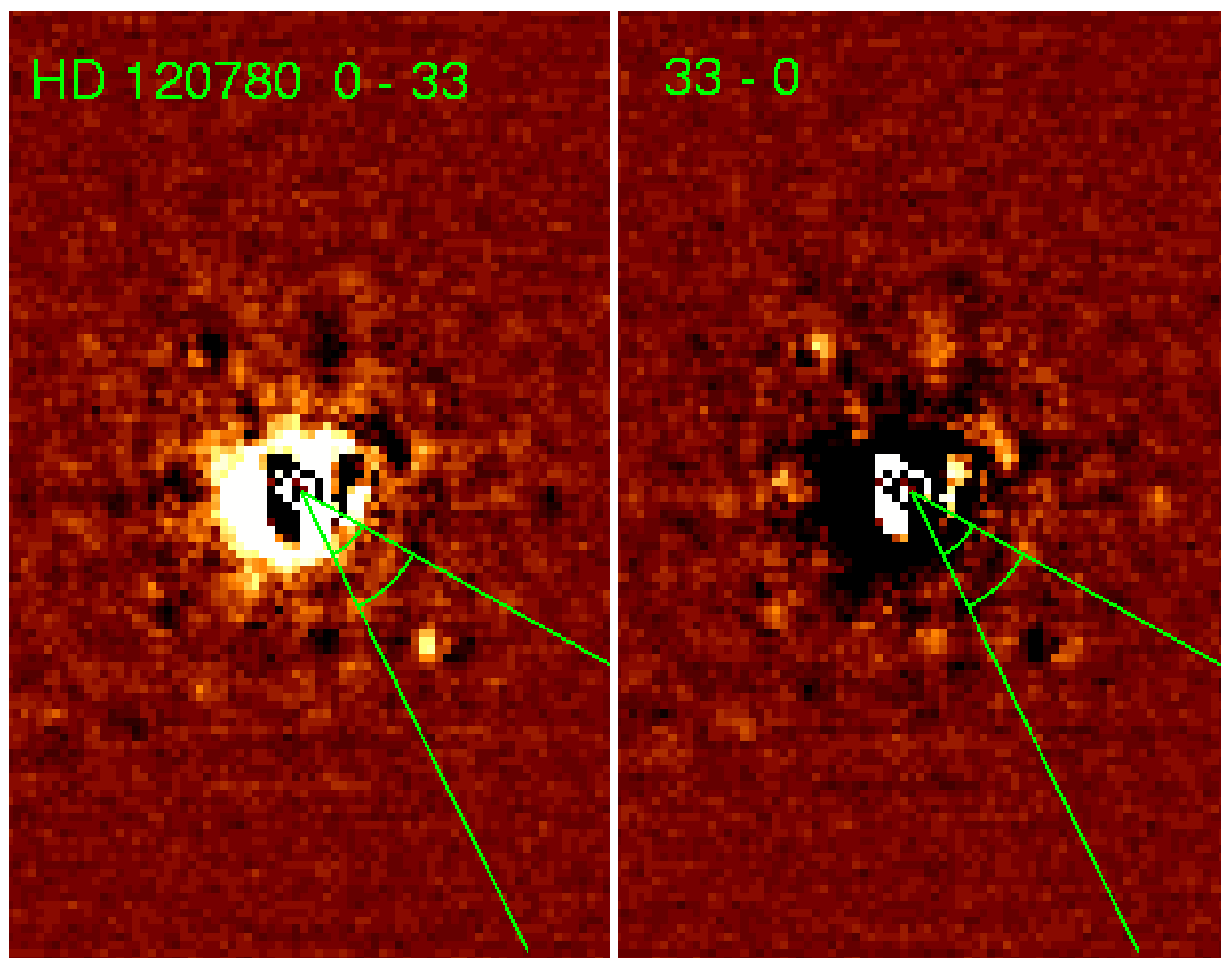}
\vspace{3.1cm}
\caption[Annotated image of HD120780]{The annotated image of the star HD120780 containing a possible detection of the companion object.  The green lines mark the projected 
roll angle of 33$^{\circ}$
of the camera to select possible candidates from speckle noise.  The positive and negative pair for this potential detection clearly intersect the green lines.  This
lends weight to the potential for this to be a single object seen through two separate roll angles of the camera.}
\label{annotate_hd120780}
\end{figure}

Figure~\ref{annotate_hd120780} shows the annotated (33$^{\circ}$ roll angle) for the star HD120780.  The angle is projected along the solid lines radiating from the 
central pixel.  The brightest object in this image is found between the projected solid lines, however since no counterpart was found this is a residual super speckle and 
shows that such artifacts still remain after SDI reduction.  The potential candidate around this star is found to reside at a separation of 0.30$\pm$0.01$''$ from the 
central pixel with a $\Delta$F1 contrast of $\sim$8.8~magnitudes, and is again well below the strong methane boundary.  Significantly though, the separation is in agreement with the 
object found around the star HD25874.  Also the position angle of the object is 240$^{\circ}$ which again is in 
agreement with the find around HD25874.  The fact that both the separation and position angle agree for the possible candidate objects around both HD25874 and HD120780 
strongly indicate that these detections are actually artifacts inherent in the reduction procedure rather than an actual detections of the companion objects to both these 
stars, possibly spider arm residuals.

\begin{figure}
\vspace{2.5cm}
\hspace{-4.0cm}
\includegraphics{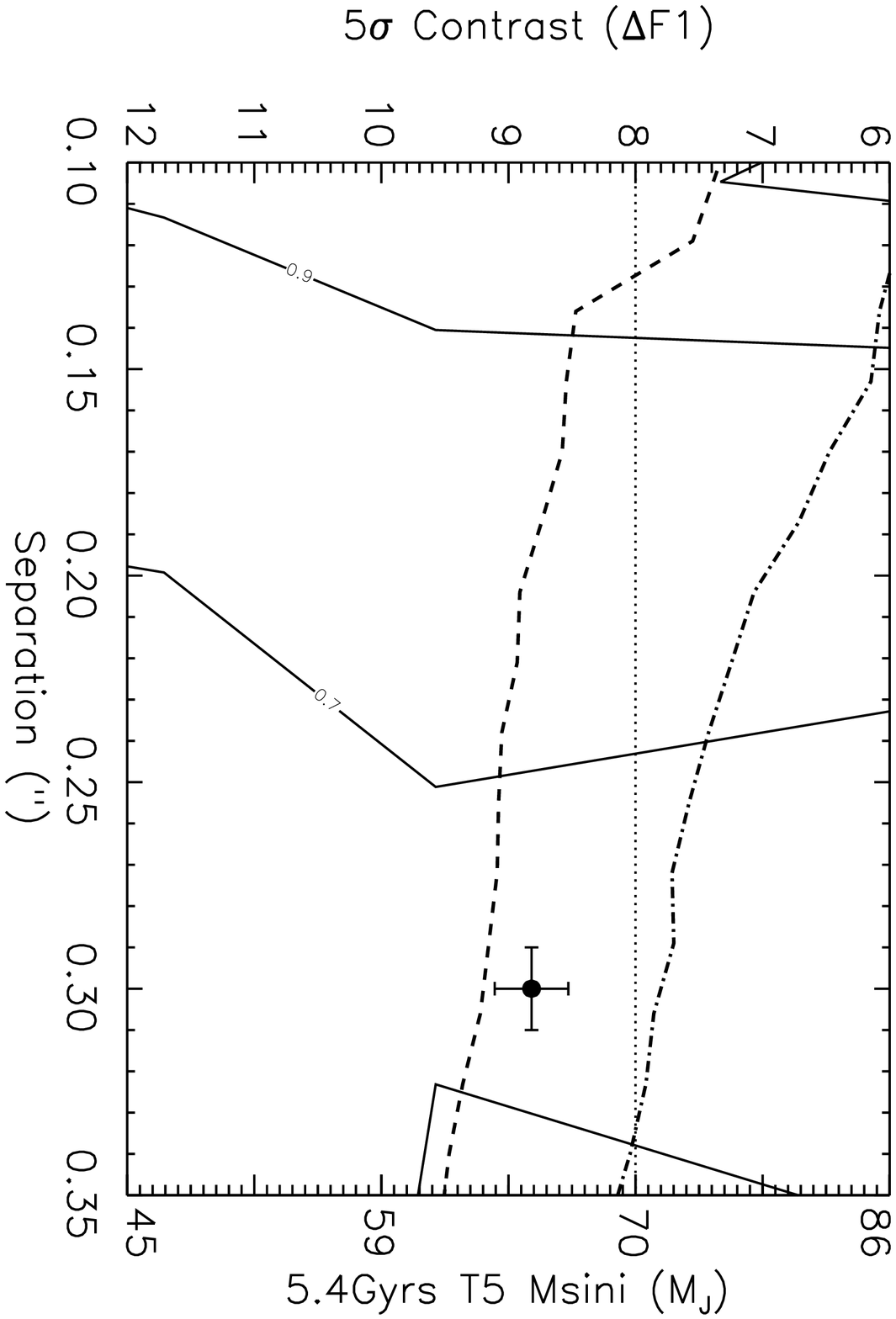}
\vspace{5.5cm}
\includegraphics{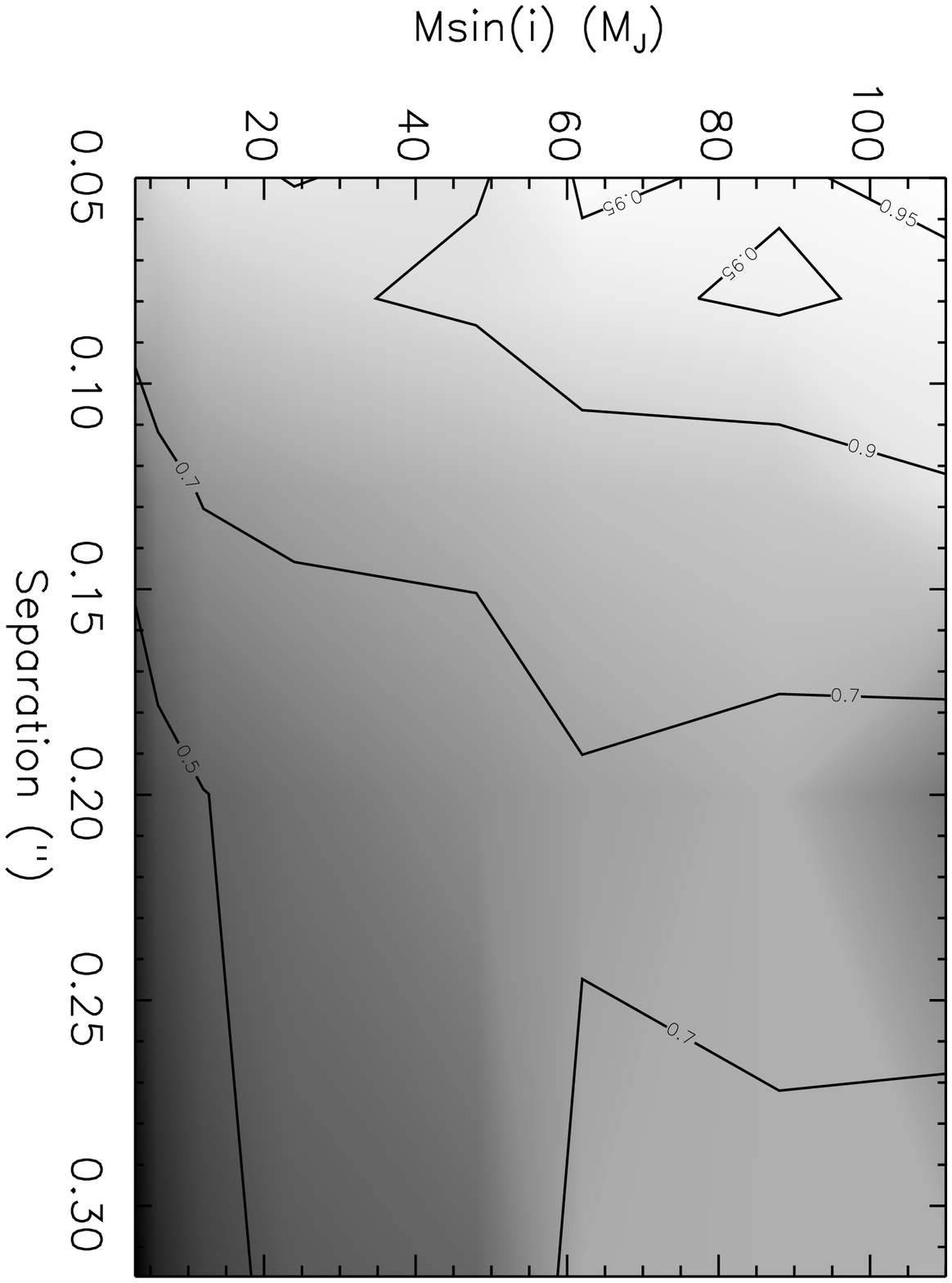}
\vspace{1.4cm}
\caption[Contrast limits for HD120780]{The upper plot shows the contrast limits ($\Delta$F1) for the star HD120780.  An SDI reduced contrast of 9.3~magnitudes is reached at 
0.3$''$, which is the separation of the potential detection (filled circle).  Note the large error bars on the contrast for this candidate since the contrasts were estimated using the 
other system fluxes since there 
were no unsaturated acquisition images for this star.  The solid curves mark the confidence limits estimated from the radial-velocity timeseries.  The right hand y-axis shows the 
expected mass for a T5 dwarf at the age assumed for these objects.  
The horizontal line represents the strong methane boundary.  The lower plot shows the radial-velocity confidence limits in mass-angular separation space.  The values inside all 
solid curves are the confidence percentages.  }
\label{hd120780_contrast}
\end{figure}

The contrast curves created for this system (Fig.~\ref{hd120780_contrast} upper) are extrapolated from the unsaturated images of the other systems.  The detection here is marked 
by the filled circle with associated uncertainties to highlight the contrasts achieved at such low angular separations.  The detectability probability in this region is found to be 50-70\%, 
again showing that more data would be useful to better constrain the companion parameters, even at such low separations.  The detectability region is better seen in
Fig.~\ref{hd120780_contrast} (lower) where again we have 
over 90\% confidence reaching out beyond 0.1$''$ from the star and with fairly high brown dwarf masses, across the stellar mass regime.  The gray scale in this region shows no real dark 
unconstrained regions, except for widely separated and low mass companions.  Below the SDI limits we see that we can 
rule out lower mass companions with 90\% confidence only out to separations of $\sim$0.14$''$ ($\sim$2.38~AU) at best.  Yet above the 1$\sigma$ confidence level (70\%) we can 
rule out a range of lower mass companions reaching as far out as 0.25$''$ (4.25~AU).  Finally no other longer period methane rich companions were detected out to a separation of 
$\sim$34AU.

\begin{table*}
\caption[All Results]{Summary of results from the combined SDI and radial-velocity constraints for all stars.} \label{tab:results}
\center
\begin{tabular}{ccccccccc}
\hline
\multicolumn{1}{c}{Star}& \multicolumn{1}{c}{AO (0.5$''$)}& \multicolumn{1}{c}{AO (1.0$''$)}& \multicolumn{1}{c}{SDI (0.5$''$)} & \multicolumn{1}{c}{SDI (1.0$''$)} & \multicolumn{1}{c}{RV 70\%}& \multicolumn{1}{c}{RV 90\%}& \multicolumn{1}{c}{RV 95\%} \\ \hline
& & & & & & & \\

HD25874   & 9.5 | 71M$_{\rm{J}}$ & 11.2 | 68M$_{\rm{J}}$ & 11.2 | 68M$_{\rm{J}}$ & 11.3 | 67M$_{\rm{J}}$ & 2M$_{\rm{J}}$ & 4M$_{\rm{J}}$ &  9M$_{\rm{J}}$ \\
HD32778   & 8.7 | 72M$_{\rm{J}}$ &   9.0 | 71M$_{\rm{J}}$ &   8.9 | 71M$_{\rm{J}}$ &   9.3 | 71M$_{\rm{J}}$ & -- & --  & --  \\
HD91204   & 9.5 | 70M$_{\rm{J}}$ & 10.1 | 68M$_{\rm{J}}$ & 10.0 | 68M$_{\rm{J}}$ & 10.0 | 67M$_{\rm{J}}$ & 3M$_{\rm{J}}$ & 13M$_{\rm{J}}$ & 30M$_{\rm{J}}$  \\
HD120780 & 8.9 | 66M$_{\rm{J}}$ &   9.8 | 60M$_{\rm{J}}$ & 10.0 | 59M$_{\rm{J}}$ & 10.3 | 57M$_{\rm{J}}$ & 3M$_{\rm{J}}$ & 35M$_{\rm{J}}$ & 23M$_{\rm{J}}$  \\
HD145825 & 9.3 | 57M$_{\rm{J}}$ & 10.9 | 41M$_{\rm{J}}$ & 10.6 | 44M$_{\rm{J}}$ & 11.0 | 40M$_{\rm{J}}$ & 3M$_{\rm{J}}$ & 11M$_{\rm{J}}$ & 15M$_{\rm{J}}$ \\

\hline
\end{tabular}

\medskip
Columns 2 and 3 show the AO reduced contrast limits and their associated T5 mass limits at the age of each star for angular separations of 0.5$''$ and 1.0$''$ respectively.  Columns 
4 and 5 show the same results but for the SDI reduction.  Columns 6--8 show the minimum mass that was ruled out at confidence levels of 70,~90~and~95\% respectively.  These 
masses were for all separations and therefore show only the lowest mass reached by the simulations and are not constrained within a separation limit.  Note the lack of this data for 
HD32778 due to the limited radial-velocity data set.

\end{table*}

\section{Summary}

We have performed a targeted direct imaging program to detect cool companions orbiting within 2$''$ of their parent stars.  The stars were drawn from the AAPS and Keck
planet search projects and consist of objects that exhibit large radial-velocity variation over several years.  These radial-velocities indicate the
presence of a massive companion on a long period orbit.  Five stars were examined with the NACO-SDI system on the VLT in Paranal, Chile.  From the five, 
two possible detections were found around the stars HD25874 and HD120780.  However, further analysis of these detections indicate they are probable residual artifacts since 
they are found to be located at the same distance from the central pixel and the same position angle in each of the images for each star.  In addition we also present detectabilities 
for each system by analysing the radial-velocity information we have acquired.  Each of these detections lie within sensitivity boundaries for these stars of between 50-70\%, 
meaning they could not be ruled out with any high degree of certainty from the radial-velocity data.   

We have summarised the results of this work in Table~\ref{tab:results}, which shows both the broad-band AO and 
narrow-band SDI reduced contrasts and mass limits at separations of 0.5$''$ and 1.0$''$.  Also the sensitivity limits from the radial-velocity data have been summarised 
at confidence limits of 70,~90~and~95\% for each system, highlighting the minimum mass possibly detectable at each confidence level.  The table shows the contrasts in 
magnitudes and mass limits in Jupiter-masses at 0.5$''$ and 1.0$''$ angular separations for 
AO and SDI images and shows the AO performs almost as well as the SDI at separations of $\ge$1.0$''$.  Also since the stars are fairly old the AO reduction tends to reach only a few 
Jupiter-masses above the SDI reduction for most of the sample, however when the star is fairly young (HD145825) the SDI reaches far deeper than AO i.e. 13M$_{\rm{J}}$ in this 
case.  Also from this table we see that the radial-velocity confidence limits are all generally the same, apart from HD32778 due its limited number of data points and temporal 
coverage.

As the stars chosen are fairly old, a 
consequence of the radial-velocity selection method, the companions must be sufficiently massive to lie within the detection threshold of the instrument.  Also since three of the five 
stars were found to have 'liner' trends, and we expect these to be sufficiently massive, then we suspect either the uncertainty in age means we are underestimating our mass thresholds, 
the models are over estimating the magnitudes of the companions, the companions are aligned such that they are found behind, or very close to the central star ($\le$0.1$''$) 
or the companions are sufficiently far out in the system that they are off the 2$''$ contrast limit.  A combination of the these mechanisms are probably at work.  We find 5$\sigma$ 
($\Delta$F1) contrasts of 11.5~magnitudes are possible using this method around bright F-K type stars (and 5$\sigma$ $H$-band contrasts of 12~magnitudes for mid T-like objects).  
Such contrasts allow access to long period, massive ($\sim\ge$40M$_{\rm{J}}$) methane objects for stars that typically constitute the bulk of radial-velocity programmes.  In the future 
similar analyses as those employed here will lead to a greater understanding of the properties of exoplanets and brown dwarfs when extreme-AO systems can gain the contrasts necessary 
to directly image known planetary-mass companions detected by ongoing Doppler programs.

\bibliographystyle{aa}
\bibliography{refs}

\begin{thebibliography}{47}
\expandafter\ifx\csname natexlab\endcsname\relax\def\natexlab#1{#1}\fi

\bibitem[{{Allard} {et~al.}(1997){Allard}, {Hauschildt}, {Alexander}, \&
  {Starrfield}}]{allard}
{Allard}, F., {Hauschildt}, P.~H., {Alexander}, D.~R., \& {Starrfield}, S.
  1997, ARA\&A, 35, 137

\bibitem[{{Baraffe} {et~al.}(2003){Baraffe}, {Chabrier}, {Barman}, {Allard}, \&
  {Hauschildt}}]{baraffe03}
{Baraffe}, I., {Chabrier}, G., {Barman}, T.~S., {Allard}, F., \& {Hauschildt},
  P.~H. 2003, A\&A, 402, 701

\bibitem[{{Biller} {et~al.}(2004){Biller}, {Close}, {Lenzen}, {Brandner},
  {McCarthy}, {Nielsen}, \& {Hartung}}]{biller05}
{Biller}, B.~A., {Close}, L., {Lenzen}, R., {et~al.} 2004, in Advancements in
  Adaptive Optics. Edited by Domenico B. Calia, Brent L. Ellerbroek, and
  Roberto Ragazzoni. Proceedings of the SPIE, Volume 5490, pp. 389-397 (2004).,
  ed. D.~{Bonaccini Calia}, B.~L. {Ellerbroek}, \& R.~{Ragazzoni}, 389--397

\bibitem[{{Biller} {et~al.}(2007){Biller}, {Close}, {Masciadri}, {Nielsen},
  {Lenzen}, {Brandner}, {McCarthy}, {Hartung}, {Kellner}, {Mamajek}, {Henning},
  {Miller}, {Kenworthy}, \& {Kulesa}}]{biller07}
{Biller}, B.~A., {Close}, L.~M., {Masciadri}, E., {et~al.} 2007, ApJS, 173, 143

\bibitem[{{Burrows} {et~al.}(1997){Burrows}, {Marley}, {Hubbard}, {Lunine},
  {Guillot}, {Saumon}, {Freedman}, {Sudarsky}, \& {Sharp}}]{burrows}
{Burrows}, A., {Marley}, M., {Hubbard}, W.~B., {et~al.} 1997, ApJ, 491, 856

\bibitem[{{Butler} {et~al.}(1996){Butler}, {Marcy}, {Williams}, {McCarthy},
  {Dosanjh}, \& {Vogt}}]{butler}
{Butler}, R.~P., {Marcy}, G.~W., {Williams}, E., {et~al.} 1996, PASP, 108, 500

\bibitem[{{Butler} {et~al.}(2001){Butler}, {Tinney}, {Marcy}, {Jones}, {Penny},
  \& {Apps}}]{butler01}
{Butler}, R.~P., {Tinney}, C.~G., {Marcy}, G.~W., {et~al.} 2001, ApJ, 555, 410

\bibitem[{{Butler} {et~al.}(2006){Butler}, {Wright}, {Marcy}, {Fischer},
  {Vogt}, {Tinney}, {Jones}, {Carter}, {Johnson}, {McCarthy}, \&
  {Penny}}]{butler06}
{Butler}, R.~P., {Wright}, J.~T., {Marcy}, G.~W., {et~al.} 2006, ApJ, 646, 505

\bibitem[{{Carpenter}(2001)}]{carpenter01}
{Carpenter}, J.~M. 2001, AJ, 121, 2851

\bibitem[{{Chauvin} {et~al.}(2004){Chauvin}, {Lagrange}, {Dumas}, {Zuckerman},
  {Mouillet}, {Song}, {Beuzit}, \& {Lowrance}}]{chauvin}
{Chauvin}, G., {Lagrange}, A.-M., {Dumas}, C., {et~al.} 2004, A\&A, 425, L29

\bibitem[{{Close} {et~al.}(2005){Close}, {Lenzen}, {Guirado}, {Nielsen},
  {Mamajek}, {Brandner}, {Hartung}, {Lidman}, \& {Biller}}]{close05}
{Close}, L.~M., {Lenzen}, R., {Guirado}, J.~C., {et~al.} 2005, Nature, 433, 286

\bibitem[{{Duquennoy} \& {Mayor}(1991{\natexlab{a}})}]{duquennoy}
{Duquennoy}, A. \& {Mayor}, M. 1991{\natexlab{a}}, A\&A, 248, 485

\bibitem[{{Duquennoy} \& {Mayor}(1991{\natexlab{b}})}]{duquennoy91}
{Duquennoy}, A. \& {Mayor}, M. 1991{\natexlab{b}}, A\&A, 248, 485

\bibitem[{{Duquennoy} {et~al.}(1992){Duquennoy}, {Mayor}, {Andersen},
  {Carquillat}, \& {North}}]{duquennoy92}
{Duquennoy}, A., {Mayor}, M., {Andersen}, J., {Carquillat}, J.~M., \& {North},
  P. 1992, A\&A, 254, L13+

\bibitem[{{Grether} \& {Lineweaver}(2006)}]{grether06}
{Grether}, D. \& {Lineweaver}, C.~H. 2006, ApJ, 640, 1051

\bibitem[{{Henry} {et~al.}(1996){Henry}, {Soderblom}, {Donahue}, \&
  {Baliunas}}]{henry}
{Henry}, T.~J., {Soderblom}, D.~R., {Donahue}, R.~A., \& {Baliunas}, S.~L.
  1996, AJ, 111, 439

\bibitem[{{Jenkins} {et~al.}(2010){Jenkins}, {Blundell}, {Jones}, {Butler},
  {McCarthy}, {Tinney}, {Marcy}, {Penny}, \& {Carter}}]{jenkins10}
{Jenkins}, J., {Blundell}, J., {Jones}, H., {et~al.} 2010, MNRAS, submitted

\bibitem[{{Jenkins} {et~al.}(2009){Jenkins}, {Jones}, {Go{\'z}dziewski},
  {Migaszewski}, {Barnes}, {Jones}, {Rojo}, {Pinfield}, {Day-Jones}, \&
  {Hoyer}}]{jenkins09a}
{Jenkins}, J.~S., {Jones}, H.~R.~A., {Go{\'z}dziewski}, K., {et~al.} 2009,
  MNRAS, 398, 911

\bibitem[{{Jenkins} {et~al.}(2008){Jenkins}, {Jones}, {Pavlenko}, {Pinfield},
  {Barnes}, \& {Lyubchik}}]{jenkins08}
{Jenkins}, J.~S., {Jones}, H.~R.~A., {Pavlenko}, Y., {et~al.} 2008, A\&A, 485,
  571

\bibitem[{{Jenkins} {et~al.}(2006){Jenkins}, {Jones}, {Tinney}, {Butler},
  {McCarthy}, {Marcy}, {Pinfield}, {Carter}, \& {Penny}}]{jenkins06c}
{Jenkins}, J.~S., {Jones}, H.~R.~A., {Tinney}, C.~G., {et~al.} 2006, MNRAS,
  372, 163

\bibitem[{{Jones} {et~al.}(2009){Jones}, {Butler}, {Tinney}, {O'Toole},
  {Wittenmyer}, {Henry}, {Meschiari}, {Vogt}, {Rivera}, {Laughlin}, {Carter},
  {Bailey}, \& {Jenkins}}]{jones10}
{Jones}, H.~R.~A., {Butler}, R.~P., {Tinney}, C.~G., {et~al.} 2009, ArXiv
  e-prints

\bibitem[{{Jones} {et~al.}(2002){Jones}, {Paul Butler}, {Tinney}, {Marcy},
  {Penny}, {McCarthy}, {Carter}, \& {Pourbaix}}]{jones02a}
{Jones}, H.~R.~A., {Paul Butler}, R., {Tinney}, C.~G., {et~al.} 2002, MNRAS,
  333, 871

\bibitem[{{Kalas} {et~al.}(2008){Kalas}, {Graham}, {Chiang}, {Fitzgerald},
  {Clampin}, {Kite}, {Stapelfeldt}, {Marois}, \& {Krist}}]{kalas08}
{Kalas}, P., {Graham}, J.~R., {Chiang}, E., {et~al.} 2008, Science, 322, 1345

\bibitem[{{Leggett} {et~al.}(2002){Leggett}, {Golimowski}, {Fan}, {Geballe},
  {Knapp}, {Brinkmann}, {Csabai}, {Gunn}, {Hawley}, {Henry}, {Hindsley},
  {Ivezi{\'c}}, {Lupton}, {Pier}, {Schneider}, {Smith}, {Strauss}, {Uomoto}, \&
  {York}}]{leggett}
{Leggett}, S.~K., {Golimowski}, D.~A., {Fan}, X., {et~al.} 2002, ApJ, 564, 452

\bibitem[{{Lenzen} {et~al.}(2004){Lenzen}, {Close}, {Brandner}, {Biller}, \&
  {Hartung}}]{lenzen04}
{Lenzen}, R., {Close}, L., {Brandner}, W., {Biller}, B., \& {Hartung}, M. 2004,
  in Ground-based Instrumentation for Astronomy. Edited by Alan F. M. Moorwood
  and Iye Masanori. Proceedings of the SPIE, Volume 5492, pp. 970-977 (2004).,
  ed. A.~F.~M. {Moorwood} \& M.~{Iye}, 970--977

\bibitem[{{Liu} {et~al.}(2002){Liu}, {Fischer}, {Graham}, {Lloyd}, {Marcy}, \&
  {Butler}}]{liu02}
{Liu}, M.~C., {Fischer}, D.~A., {Graham}, J.~R., {et~al.} 2002, ApJ, 571, 519

\bibitem[{{Marcy} \& {Butler}(1992)}]{marcy92}
{Marcy}, G.~W. \& {Butler}, R.~P. 1992, PASP, 104, 270

\bibitem[{{Marcy} \& {Butler}(2000)}]{marcy}
{Marcy}, G.~W. \& {Butler}, R.~P. 2000, PASP, 112, 137

\bibitem[{{Marcy} {et~al.}(2005){Marcy}, {Butler}, {Vogt}, {Fischer}, {Henry},
  {Laughlin}, {Wright}, \& {Johnson}}]{marcy05a}
{Marcy}, G.~W., {Butler}, R.~P., {Vogt}, S.~S., {et~al.} 2005, ApJ, 619, 570

\bibitem[{{Marois} {et~al.}(2008){Marois}, {Macintosh}, {Barman}, {Zuckerman},
  {Song}, {Patience}, {Lafreni{\`e}re}, \& {Doyon}}]{marois08}
{Marois}, C., {Macintosh}, B., {Barman}, T., {et~al.} 2008, Science, 322, 1348

\bibitem[{{McCarthy} \& {Zuckerman}(2004)}]{mccarthy04}
{McCarthy}, C. \& {Zuckerman}, B. 2004, AJ, 127, 2871

\bibitem[{{McCaughrean}(2003)}]{mccaughrean03}
{McCaughrean}, M.~J. 2003, American Astronomical Society Meeting Abstracts,
  203, 124.04

\bibitem[{{Meschiari} {et~al.}(2009){Meschiari}, {Wolf}, {Rivera}, {Laughlin},
  {Vogt}, \& {Butler}}]{meschiari09}
{Meschiari}, S., {Wolf}, A.~S., {Rivera}, E., {et~al.} 2009, PASP, 121, 1016

\bibitem[{{Mugrauer} \& {Neuh{\"a}user}(2005)}]{mugrauer}
{Mugrauer}, M. \& {Neuh{\"a}user}, R. 2005, MNRAS, 361, L15

\bibitem[{{Mugrauer} {et~al.}(2006){Mugrauer}, {Neuh{\"a}user}, {Mazeh},
  {Guenther}, {Fern{\'a}ndez}, \& {Broeg}}]{mugrauer06}
{Mugrauer}, M., {Neuh{\"a}user}, R., {Mazeh}, T., {et~al.} 2006, Astronomische
  Nachrichten, 327, 321

\bibitem[{{Mugrauer} {et~al.}(2007){Mugrauer}, {Seifahrt}, \&
  {Neuh{\"a}user}}]{mugrauer07}
{Mugrauer}, M., {Seifahrt}, A., \& {Neuh{\"a}user}, R. 2007, MNRAS, 378, 1328

\bibitem[{{Neuh{\"a}user} {et~al.}(2005){Neuh{\"a}user}, {Guenther},
  {Wuchterl}, {Mugrauer}, {Bedalov}, \& {Hauschildt}}]{neuhauser}
{Neuh{\"a}user}, R., {Guenther}, E.~W., {Wuchterl}, G., {et~al.} 2005, A\&A,
  435, L13

\bibitem[{{Nidever} {et~al.}(2002){Nidever}, {Marcy}, {Butler}, {Fischer}, \&
  {Vogt}}]{nidever}
{Nidever}, D.~L., {Marcy}, G.~W., {Butler}, R.~P., {Fischer}, D.~A., \& {Vogt},
  S.~S. 2002, ApJS, 141, 503

\bibitem[{{O'Toole} {et~al.}(2009){O'Toole}, {Tinney}, {Jones}, {Butler},
  {Marcy}, {Carter}, \& {Bailey}}]{otoole09}
{O'Toole}, S.~J., {Tinney}, C.~G., {Jones}, H.~R.~A., {et~al.} 2009, MNRAS,
  392, 641

\bibitem[{{Rousset} {et~al.}(2003){Rousset}, {Lacombe}, {Puget}, {Hubin},
  {Gendron}, {Fusco}, {Arsenault}, {Charton}, {Feautrier}, {Gigan}, {Kern},
  {Lagrange}, {Madec}, {Mouillet}, {Rabaud}, {Rabou}, {Stadler}, \&
  {Zins}}]{rousset03}
{Rousset}, G., {Lacombe}, F., {Puget}, P., {et~al.} 2003, in Adaptive Optical
  System Technologies II. Edited by Wizinowich, Peter L.; Bonaccini, Domenico.
  Proceedings of the SPIE, Volume 4839, pp. 140-149 (2003)., ed. P.~L.
  {Wizinowich} \& D.~{Bonaccini}, 140--149

\bibitem[{{Takeda} {et~al.}(2007){Takeda}, {Ford}, {Sills}, {Rasio}, {Fischer},
  \& {Valenti}}]{takeda07}
{Takeda}, G., {Ford}, E.~B., {Sills}, A., {et~al.} 2007, ApJS, 168, 297

\bibitem[{{Valenti} \& {Fischer}(2005)}]{valenti05}
{Valenti}, J.~A. \& {Fischer}, D.~A. 2005, ApJS, 159, 141

\bibitem[{{van Leeuwen} \& {Fantino}(2005)}]{vanleeuwen05}
{van Leeuwen}, F. \& {Fantino}, E. 2005, A\&A, 439, 791

\bibitem[{{Wittenmyer} {et~al.}(2009){Wittenmyer}, {Endl}, {Cochran},
  {Ram{\'{\i}}rez}, {Reffert}, {MacQueen}, \& {Shetrone}}]{wittenmyer09}
{Wittenmyer}, R.~A., {Endl}, M., {Cochran}, W.~D., {et~al.} 2009, AJ, 137, 3529

\bibitem[{{Wright}(2005)}]{wright05}
{Wright}, J.~T. 2005, PASP, 117, 657

\bibitem[{{Wright} {et~al.}(2007){Wright}, {Marcy}, {Fischer}, {Butler},
  {Vogt}, {Tinney}, {Jones}, {Carter}, {Johnson}, {McCarthy}, \&
  {Apps}}]{wright07}
{Wright}, J.~T., {Marcy}, G.~W., {Fischer}, D.~A., {et~al.} 2007, ApJ, 657, 533

\bibitem[{{Wright} {et~al.}(2009){Wright}, {Upadhyay}, {Marcy}, {Fischer},
  {Ford}, \& {Johnson}}]{wright09}
{Wright}, J.~T., {Upadhyay}, S., {Marcy}, G.~W., {et~al.} 2009, ApJ, 693, 1084

\end{thebibliography}

\end{document}